\documentclass[aps,prc,superscriptaddress,twoside,twocolumn,nofootinbib,10pt,showpacs,floatfix]{revtex4-1}

\usepackage{braket}
\usepackage{bm}
\usepackage{amsmath}
\usepackage[pdftex]{graphicx}
\usepackage{multirow}
\newcommand{\Slash}[1]{{\ooalign{\hfil/\hfil\crcr$#1$}}}
\usepackage{colortbl}
\usepackage{here}

\usepackage{ulem} 
\usepackage{amssymb}

\begin{document}

\title{Chiral partner structure of light nucleons in an extended parity doublet model}

\author{Takahiro Yamazaki}
\email{yamazaki@hken.phys.nagoya-u.ac.jp}
\author{Masayasu Harada}
\email{harada@hken.phys.nagoya-u.ac.jp}
\affiliation{Department of Physics,  Nagoya University, Nagoya, 464-8602, Japan}
\date{\today}

\newcommand\sect[1]{\emph{#1}---}
\begin{abstract}
We study chiral partner structure of four light nucleons, $N(939)$, $N(1440)$, $N(1535)$ and $N(1650)$ using an effective chiral model based on the parity doublet structure.
In our model we introduce four chiral representations, 
$({\bf 1},{\bf 2})$, $({\bf 2},{\bf 1})$, $({\bf 2},{\bf 3})$ and $({\bf 3},{\bf 2})$
under ${\rm SU}(2)_{\rm L} \otimes {\rm SU}(2)_{\rm R}$ symmetry.
We determine the model parameters by fitting them to available experimental values of masses, widths and the axial charge of $N(939)$ together with the axial charges of $N(1535)$ and $N(1650)$ by lattice analyses.
We find five groups of solutions: In a group
the chiral partner to $N(939)$ is $N(1440)$ having small chiral invariant mass. In another group, the chiral partner is a mixture of $N(1535)$ and $N(1650)$ having a large chiral invariant mass.
We claim that off-diagonal elements of axial-charge matrix can be used for distinguishing these groups.
We also discuss changes of masses associated with chiral symmetry restoration, which could emerge in high density matter.  
\end{abstract}
\maketitle
\section{Introduction}
\label{sec:Intro}
One of the most important features of QCD relevant to the low-energy hadron physics is the chiral symmetry and its spontanious breaking. The spontanious symmetry breaking generates mass differences between chiral partners 
as well as the mixing among different chiral representations. It is interesting  to study the role of the chiral symmetry breaking to determine the properties and structures of baryons
such as 
amount of the masses of baryons 
generated by the chiral symmetry breaking and
the chiral partner of ground state nucleon.

In a hadronic model for light nucleons based on the parity doublet structure~\cite{DeTar,Jido1,Jido2,Gallas:2009qp,Gallas} 
the chiral partner of $N(939)$. 
One of the important feature of the model 
is the existence of the chiral invariant mass denoted by $m_0$ which is not originated from the
spontaneous chiral symmetry breaking.
In other word,
the mass splitting between  $N(939)$ and its chiral partner $N(1535)$
 denoted by $m_0$, which implies that the masses of $N(939)$ and its chiral partner tend to $m_0$ when the chiral symmetry is restored and their mass splitting is given by spontanious chiral symmetry braking. 

This parity doublet structure is extended to include hyperons and/or excited nucleons such as $N(1440)$ and $N(1650)$ (see, e.g. Refs.~\cite{Nemoto1,Chen:2008qv,%
Dmitrasinovic:2009vp,Dmitrasinovic:2009vy,Chen:2009sf,Chen:2010ba,Chen:2011rh,
Dmitrasinovic1,Nishihara:2015fka,Olbrich:2015gln,Olbrich:2017fsd}).

Particularly, 
in Ref.~\cite{Olbrich:2015gln}, two $[(\bf{1},\bf{2})\oplus(\bf{2},\bf{1})]$ representations under the chiral $\mbox{SU}(2)_L \otimes \mbox{SU}(2)_R$ are introduced to study $N(939)$, $N(1440)$, $N(1535)$ and $N(1650)$.
Two types of  solutions were found:  In one type, the chiral partner of $N(939)$ is a mixture of $N(1535)$ and $N(1650)$ and two chiral invariant masses are almost $200$\,MeV. 
In another type, $N(939)$ and  $N(1535)$ are chiral partners to each other and  two chiral invariant masses are about $1000$\,MeV. 
On the other hand, in Ref.~\cite{Nishihara:2015fka}, 
a model is constructed in the three-flavor framework to further introduce the chiral $[(\bf{3},\bf{6})\oplus(\bf{6},\bf{3})]$ representations under the chiral $\mbox{SU}(3)_L \otimes \mbox{SU}(3)_R$ symmetry, which correspond to the chiral  $[(\bf{2},\bf{3})\oplus(\bf{3},\bf{2})]$ representations under the chiral $\mbox{SU}(2)_L \otimes \mbox{SU}(2)_R$ symmetry.
It was shown that the $N(939)$ is dominated by the $[(\bf{3},\bf{6})\oplus(\bf{6},\bf{3})]$
to reproduce the axial charge of nucleon as pointed in Ref.~\cite{Dmitrasinovic:2009vp}, and that the chiral invariant mass of the $N(939)$ is about $800$\,MeV.
The large value of the chiral invariant mass of $N(939)$ seems consistent with the 
results by the lattice QCD analysis in  Ref.~\cite{Aarts:2015mma,Aarts:2017rrl},
which shows that
is almost constant even if temperature is increased. 

In this paper, we make a general analysis using a two-flavor parity doublet model including chiral $[(\bf{1},\bf{2})\oplus(\bf{2},\bf{1})]$ and $[(\bf{2},\bf{3})\oplus(\bf{3},\bf{2})]$ representations 
under the chiral $\mbox{SU}(2)_L \otimes \mbox{SU}(2)_R$ symmetry, with including 
derivative 
interactions to the pion fields. 
We will show that there exist five groups of solutions 
distinguishable 
by chiral inavarinat masses and mixing rates of nucleons. 
In a group of solutions, the 
chiral partner of $N(939)$ is $N(1440)$ having small chiral invariant mass of 
about 100MeV and $N(939)$ is dominated by $[(\bf{2},\bf{3})\oplus(\bf{3},\bf{2})]$ representation. 
In another group, on the other hand, $N(939)$ belongs dominantly $[(\bf{1},\bf{2})\oplus(\bf{2},\bf{1})]$ representation having a large chiral invariant mass, and chiral partner of $N(939)$ is a mixture of $N(1535)$ and $N(1650)$. 
Futhermore, we give predictions of off-diagonal elements of axial-charge matrix, which could be checked in future lattice analysis.
We also 
show changes of nucleon masses when the 
vacuum expectation value of $\sigma$, which is an order parameter of the spnoaneous chiral symmetry breaking, is changed. 

This paper is organized 
as follows: 
In section \ref{sec:EPDM}, 
we construct an extended model with parity doublet structure.  
Section~\ref{sec:CIM} 
is a main part, where 
we show the numerical results of fitting on the
chiral invariant masses and chiral partner structure.
In sections \ref{sec:axial charge} and \ref{sec:masses}, we study
off-diagonal components of axial-charges matrix and change of nucleons mass as predictions. 
Finally we will give a brief summary and discussions in section~\ref{sec:summary}.

\section{An Extended parity doublet model}
\label{sec:EPDM}

In this section we introduce four baryon fields with parity doublet structure and construct a Lagrangian for baryons and scalar and pseudoscalar mesons based on the ${\rm SU(2)_L}$$\otimes$${\rm SU(2)_R}$ chiral symmetry.

\subsection{Model construction}

The 
chiral representations of quarks under ${\rm SU(2)_L}$$\otimes$${\rm SU(2)_R}$ are written as
\begin{equation}
q_L\sim({\bf 2},{\bf 1})\hspace{5mm},\hspace{5mm} q_R\sim({\bf 1},{\bf 2}) \ , 
\end{equation}
where the $\bf{2}$ and $\bf{1}$ in above bracket express doublet and singlet, respectively.
Since baryons are expressed as direct products of three quarks, we have following possibilities for the chiral representations of baryons: 
\begin{align}
q\otimes q\otimes q \sim & [({\bf 2},{\bf 1})\oplus({\bf 1},{\bf 2})]^3
\nonumber
\\
\sim& 5[({\bf 2},{\bf 1})\oplus({\bf 1},{\bf 2})]
\oplus3[({\bf 3},{\bf 2})\oplus({\bf 2},{\bf 3})]
\notag \\ & 
\oplus[({\bf 4},{\bf 1})\oplus({\bf 1},{\bf 4})]
\ .
\end{align}
After the chiral symmetry is spontaneously broken down to the flavor symmetry, nucleons appear from the representations of $(\bf{2},\bf{1})\oplus(\bf{1},\bf{2})$ and $(\bf{3},\bf{2})\oplus(\bf{2},\bf{3})$. 
In this paper we introduce two baryon fields corresponding to these two representations:
\begin{eqnarray}
&\psi_{1l}\sim({\bf 2},{\bf 1}),\hspace{10mm}\psi_{1r}\sim({\bf 1},{\bf 2})&
\nonumber
\\
&\psi_{2l}\sim({\bf 1},{\bf 2}),\hspace{10mm}\psi_{2r}\sim({\bf 2},{\bf 1})& 
\nonumber
\\
&\eta_{1l}\sim({\bf 2},{\bf 3}),\hspace{10mm}\eta_{1r}\sim({\bf 3},{\bf 2})&
\nonumber
\\
&\eta_{2l}\sim({\bf 3},{\bf 2}),\hspace{10mm}\eta_{2r}\sim({\bf 2},{\bf 3})&
\label{assignment}
\end{eqnarray}
Here the subscripts $l$ and $r$ express the chirality:
\begin{eqnarray}
&\gamma_5\psi_{il}=-\psi_{il},\hspace{10mm}\gamma_5\psi_{ir}=+\psi_{ir}&
\nonumber
\\
&\gamma_5\eta_{il}=-\eta_{il},\hspace{10mm}\gamma_5\eta_{ir}=+\eta_{ir}&
\end{eqnarray}
for $i=1,2$.

For clarifying the representations under the chiral symmetry, 
we explicitly write the superscripts of the baryon fields as
\begin{align}
&(\psi_{1l})^a , (\psi_{1r})^\alpha , (\psi_{2l})^\alpha , (\psi_{2r})^a
\nonumber
\\
&\eta_{1l}^{(a,\alpha\beta)},\eta_{1r}^{(ab,\alpha)},\eta_{2l}^{(ab,\alpha)},\eta_{2r}^{(a,\alpha\beta)}
\end{align}
where $a,b=1,2$ are for ${\rm SU(2)_L}$ and $\alpha , \beta=1,2$ for ${\rm SU(2)_R}$. 
Note that the superscripts $ab$ and $\alpha \beta$ of $\eta$ fields are symmetrized to express ${\bf 3}$ representation: e.g.
$\eta_{1l}^{(a,\alpha\beta)} = \eta_{1l}^{(a,\beta\alpha)}$.  
The transformation properties under the parity and the charge conjugation are defined as
\begin{align}
&\Psi_{1l,1r}\ \mathop{\to}_{P}\ \gamma_0\Psi_{1r,1l}\hspace{5mm},\hspace{5mm}\Psi_{2l,2r} \ \mathop{\to}_P\  -\gamma_0\Psi_{2r,2l}\hspace{5mm},\hspace{5mm}
\\
&\Psi_{1l,1r}\ \mathop{\to}_{C}\ C(\bar{\Psi}_{1r,1l})^T\ ,\hspace{5mm}\Psi_{2l,2r}\ \mathop{\to}_{C}\  -C(\bar{\Psi}_{2r,2l})^T\ ,\hspace{5mm}
\end{align}
where
$C=i\gamma^2\gamma^0$ and $\Psi=\psi,\eta$.
The covariant derivatives for the fields are expressed as
\begin{eqnarray}
D_{\mu}\psi_{1r,2l}&=&(\partial_{\mu}-i\mathcal{R}_{\mu})\psi_{1r,2l}  \ ,
\nonumber
\\
D_{\mu}\psi_{1l,2r}&=&(\partial_{\mu}-i\mathcal{L}_{\mu})\psi_{1l,2r} \ ,
\end{eqnarray}
and 
\begin{eqnarray}
(D_{\mu}\eta_{1l,2r})^{(a,\alpha\beta)}&=&\partial_{\mu}\eta_{1l,2r}^{(a,\alpha\beta)}-i(\mathcal{L}_{\mu})^a_b\eta_{1l,2r}^{(b,\alpha\beta)} \nonumber \\
&&-i[(\mathcal{R}_{\mu})^{\alpha}_{\rho}\delta_{\sigma}^{\beta}+\delta_{\rho}^{\alpha}(\mathcal{R}_{\mu})^{\beta}_{\sigma}]\eta_{1l,2r}^{(a,\rho\sigma)} \ ,
\nonumber
\\
(D_{\mu}\eta_{1r,2l})^{(ab,\alpha)}&=&\partial_{\mu}\eta_{1r,2l}^{(ab,\alpha)}-i(\mathcal{R}_{\mu})^{\alpha}_{\beta}\eta_{1r,2l}^{(ab,\beta)} \nonumber \\
&&-i[(\mathcal{L}_{\mu})^{a}_{c}\delta_{d}^{b}+\delta_{c}^{a}(\mathcal{L}_{\mu})^{b}_{d}]\eta_{1r,2l}^{(cd,a)} \ ,
\end{eqnarray}
where $\mathcal{L}_\mu$ and $\mathcal{R}_\mu$ are the external gauge fields introduced by gauging the chiral ${\rm SU(2)_L}$$\otimes$${\rm SU(2)_R}$ symmetry.

Next we introduce a 2$\times$2 matrix field $M$ expressing a nonet of scalar and pseudoscalar mesons made of a quark and an antiquark.
The representation under ${\rm SU(2)_L}$$\otimes$${\rm SU(2)_R}$ of the $M$ is
\begin{equation}
M=\frac{\sigma}{2}+i{\vec \pi} \cdot {\vec T}\sim({\bf 2},{\bf 2}).
\end{equation}
The transformation properties under the parity and the charge congugation are given by  
\begin{align}
M\ \mathop{\to}_P\ M^{\dag} \ , \quad
M\ \mathop{\to}_{C}\ M^{T} \ .
\end{align}
The covariant derivative for $M$ is expressed as
\begin{equation}
(D_{\mu}M)^a_\alpha=\partial_{\mu}M^a_{\alpha}-i(\mathcal{L}_{\mu})^a_bM^b_\alpha+iM_\beta^a(\mathcal{R}_{\mu})_{\alpha}^\beta
\end{equation}

Using the fields  introduced above we construct a Lagrangian invariant under the chiral ${\rm SU(2)_L}$$\otimes$${\rm SU(2)_R}$ symmetry.
\begin{widetext}
Let us first consider terms including only $\psi_1$ and $\psi_2$ and their Yukawa interaction to $M$ field.  In the present analysis, we include interaction terms with one $M$ field.  Then, possible terms are expressed as 
\begin{eqnarray}
\mathcal{L}^{(1)}=&&\bar{\psi}_{1l}i\Slash{D}\psi_{1l}+\bar{\psi}_{1r}i\Slash{D}\psi_{1r} 	\nonumber 
			+\bar{\psi}_{2l}i\Slash{D}\psi_{2l}+\bar{\psi}_{2r}i\Slash{D}\psi_{2r}	
\nonumber
\\
			&&-g_1(\bar{\psi}_{1l}M\psi_{1r}+\bar{\psi}_{1r}M^{\dag}\psi_{1l})
				-g_2(\bar{\psi}_{2r}M\psi_{2l}+\bar{\psi}_{2l}M^{\dag}\psi_{2r}) 	
\nonumber
\\
			&&-m_0^{(1)}(\bar{\psi}_{1l}\psi_{2r}-\bar{\psi}_{1r}\psi_{2l}-\bar{\psi}_{2l}\psi_{1r}+\bar{\psi}_{2r}\psi_{1l}) 
\ .
\end{eqnarray}
A part including $\eta_1$ and $\eta_2$ with $M$ is 
\begin{eqnarray}
\mathcal{L}^{(2)}=&&(\bar{\eta}_{1l})_{(a,\alpha\beta)}i\Slash{D}(\eta_{1l})^{(a,\alpha\beta)}
			+(\bar{\eta}_{1r})_{(ab,\alpha)}i\Slash{D}(\eta_{1r})^{(ab,\alpha)} 
			+(\bar{\eta}_{2l})_{(ab,\alpha)}i\Slash{D}(\eta_{2l})^{(ab,\alpha)}
			+(\bar{\eta}_{2r})_{(a,\alpha\beta)}i\Slash{D}(\eta_{2r})^{(a,\alpha\beta)} 	\nonumber 
\\
			&&-g_3[(\bar{\eta}_{1r})_{(ab,\alpha)}(M)^a_{\beta}(\eta_{1l})^{(b,\alpha\beta)}
			+(\bar{\eta}_{1l})_{(a,\alpha\beta)}(M^{\dag})^{\alpha}_{b}(\eta_{1r})^{(ab,\beta)}] 	
\nonumber
\\
			&&-g_4[(\bar{\eta}_{2l})_{(ab,\alpha)}(M)^a_{\beta}(\eta_{2r})^{(b,\alpha\beta)}
			+(\bar{\eta}_{2r})_{(a,\alpha\beta)}(M^{\dag})^{\alpha}_{b}(\eta_{2l})^{(ab,\beta)}] 	\nonumber 
\\
			&&-m_0^{(2)}[(\bar{\eta}_{1l})_{(a,\alpha\beta)}(\eta_{2r})^{(a,\alpha\beta)}
			-(\bar{\eta}_{1r})_{(ab,\alpha)}(\eta_{2l})^{(ab,\alpha)}
			+(\bar{\eta}_{2r})_{(a,\alpha\beta)}(\eta_{1l})^{(a,\alpha\beta)}
			-(\bar{\eta}_{2l})_{(ab,\alpha)}(\eta_{1r})^{(ab,\alpha)}]
\ .
\end{eqnarray}
Yukawa interaction terms connecting $\psi$ fields to $\eta$ fields are expressed as
\begin{eqnarray}
\mathcal{L}^{(3)}=&&-y_1[\epsilon_{bc}(\bar{\psi}_{1r})_{\alpha}(M)^b_{\beta}(\eta_{1l})^{(c,\alpha\beta)}
			+\epsilon_{\beta\sigma}(\bar{\psi}_{1l})_a(M^{\dag})_b^{\beta}(\eta_{1r})^{(ab,\sigma)} 
			+\epsilon^{bc}(\bar{\eta}_{1l})_{(c,\alpha\beta)}(M^{\dag})^{\beta}_b(\psi_{1r})^{\alpha}
			+\epsilon^{\beta\sigma}(\bar{\eta}_{1r})_{ab,\sigma}(M)^b_{\beta}(\psi_{1l})^a] \nonumber 
\\
			&&-y_2[\epsilon_{\beta\sigma}(\bar{\psi}_{2r})_a(M^{\dag})_b^{\beta}(\eta_{2l})^{(ab,\sigma)}
			+\epsilon_{bc}(\bar{\psi}_{2l})_{\alpha}(M)^b_{\beta}(\eta_{2r})^{(c,\alpha\beta)} 
			+\epsilon^{\beta\sigma}(\bar{\eta}_{2l})_{(ab,\sigma)}(M)^b_{\beta}(\psi_{2r})^a
			+\epsilon^{bc}(\bar{\eta}_{2r})_{(c,\alpha\beta)}(M^{\dag})^{\beta}_b(\psi_{2l})^{\alpha}] \ .
\nonumber
\\
\end{eqnarray}
In addition to the non-derivative interactions shown above, we need to include derivative interactions.  Possible interaction terms including one derivative are given by 
\begin{eqnarray}
\mathcal{L}^{(4)}=\frac{i}{f_\pi}(&&-a_1[\bar{\psi}_{1l}\Slash{D}M\psi_{2l}-\bar{\psi}_{1r}\Slash{D}M^{\dag}\psi_{2r}
					+\bar{\psi}_{2l}\Slash{D}M^{\dag}\psi_{1l}-\bar{\psi}_{2r}\Slash{D}M\psi_{1r}]
\nonumber
\\
			&&-a_2[\bar{\eta}_{1r (ab,\alpha)}(\Slash{D}M)^a_{\beta}\eta_{2r}^{(b,\alpha\beta)}
-\bar{\eta}_{1l (a,\alpha\beta)}(\Slash{D}M^{\dag})^{\alpha}_{b}\eta_{2r}^{(ab,\beta)}
+\bar{\eta}_{2r(a,\alpha\beta)}(\Slash{D}M^{\dag})^{\alpha}_{b}\eta_{1r}^{(ab,\beta)}
-\bar{\eta}_{2l(ab,\alpha)}(\Slash{D}M)^a_{\beta}\eta_{1l}^{(b,\alpha\beta)}]
\nonumber
\\
&&-a_3[\epsilon_{ab}\bar{\psi}_{1r \alpha}(\Slash{D}M)^a_{\beta}\eta_{2r}^{(b,\alpha \beta)}
-\epsilon_{\alpha \beta}\bar{\psi}_{1l a}(\Slash{D}M^{\dag})^{\alpha}_b\eta_{2l}^{(ab,\beta}
+\epsilon^{ab}\bar{\eta}_{2r (b,\alpha \beta)}(\Slash{D}M^{\dag})^{\beta}_a\psi_{1r}^\alpha
-\epsilon^{\alpha \beta}\bar{\eta}_{2l (ab,\beta)}(\Slash{D}M)^b_\alpha\psi_{1l}^a]
\nonumber
\\
&&-a_4[\epsilon_{\alpha \beta}\bar{\psi}_{2r a}(\Slash{D}M^{\dag})^{\alpha}_b\eta_{1r}^{(ab,\beta)}
-\epsilon_{ab}\bar{\psi}_{2l \alpha}(\Slash{D}M)^{a}_{\beta}\eta_{1l}^{(\alpha \beta,b)}
+\epsilon^{\alpha \beta}\bar{\eta}_{1r (ab,\beta)}(\Slash{D}M)^{b}_{\alpha}\psi_{2r}^a
		-\epsilon^{ab}\bar{\eta}_{1l (\alpha \beta,b)}(\Slash{D}M^{\dag})^{\beta}_a\psi_{2l}^\alpha]) 
\nonumber
\\
\end{eqnarray}
\end{widetext}
Combining the above terms together, the Lagrangian in the present analysis is given by
\begin{equation}
\mathcal{L}=\mathcal{L}^{(1)}+\mathcal{L}^{(2)}+\mathcal{L}^{(3)}+\mathcal{L}^{(4)}+\mathcal{L}_{\rm meson} \ ,
\end{equation}
where the mesonic part $\mathcal{L}_{\rm meson}$ is written as 
\begin{equation}
\mathcal{L}_{\rm meson}={\rm Tr}[D_{\mu}M\cdot D^{\mu}M^{\dag}]-V(M)
\end{equation}
where $V(M)$ is a meson potential term.
In this paper, we do not specify the form of the potential, but we assume that this potential provides the vacuum expectation value (VEV) of $M$ as
$\langle M \rangle = \mbox{diag}( f_\pi/2 , f_\pi/2)$, where $f_\pi$ is the pion decay constant.

\subsection{Mass matrix}
We have constructed the Lagrangian by requiring the chiral ${\rm SU(2)_L}$$\otimes$${\rm SU(2)_R}$ invariance. To study the properties of nucleons, we decompose baryons in the chiral representation to irreducible representations of the flavor symmetry as
\begin{align}
& \psi_{1l,2r}=N^{(1)}_{1l,2r}\ ,\hspace{5mm}\psi_{1r,2l}=N^{(1)}_{1r,2l}\ ,				\\
& \eta_{1l,2r}^{(a,\alpha\beta)}=\Delta_{1l,2r}^{a \alpha \beta}+\frac{1}{\sqrt{6}}(\epsilon^{\alpha a}\delta^{\beta}_{k}
+\epsilon^{\beta a}\delta^{\alpha}_{k})(N^{(2)}_{1l,2r})^k \ , \\
& \eta_{1r,2l}^{(ab,\alpha)}=\Delta_{1r,2l}^{ab\alpha}+\frac{1}{\sqrt{6}}(\epsilon^{a\alpha }\delta^{b}_{k}
+\epsilon^{b\alpha }\delta^{a}_{k})(N^{(2)}_{1r,2l})^k \ .
\end{align}
In the following, for convenience we redefine the nucleon fields as:
\begin{equation}
N^{\prime(i)}_1 =N_1^{(i)} \ , \quad
N^{\prime(i)}_2 = \gamma_5 N_2^{(i)} \ , \quad ( i = 1,2) \ .
\label{redef}
\end{equation}
The mass terms for the redefined fields are expressed as
\begin{equation}
-\bar{N'}M'_NN'
\end{equation}
with
\begin{equation}
N'^T\equiv(N'^{(1)}_1\hspace{5mm}N'^{(2)}_1\hspace{5mm}N'^{(1)}_2\hspace{5mm}N'^{(2)}_2)
\end{equation}
and 
\begin{equation}
M'_N=
\begin{pmatrix}
				\frac{g_1}{2}f_\pi &-\frac{3y_1}{2\sqrt{6}}f_\pi &m^{(1)}_0 & 0\\
				-\frac{3y_1}{2\sqrt{6}}f_\pi & \frac{g_3}{4}f_\pi & 0 & m^{(2)}_0\\
				m^{(1)}_0 & 0 & -\frac{g_2}{2}f_\pi & \frac{3y_2}{2\sqrt{6}}f_\pi\\
				0 & m^{(2)}_0 & \frac{3y_2}{2\sqrt{6}}f_\pi & -\frac{g_4}{4}f_\pi
\end{pmatrix}.
\end{equation}
The mass eigenstates denoted by
\begin{equation}
N'^T_{\rm phys}\equiv(N'^{(1)}_+\hspace{5mm}N'^{(2)}_+\hspace{5mm}N'^{(1)}_-\hspace{5mm}N'^{(2)}_-)_{\rm phys}
\end{equation}
are obtained by diagonalizing the above mass matrix $M'_N$. 
We note that parities of all fields in $N'$ are even due to the redefinition given in Eq.~(\ref{redef}).  Two eigenvalues of the mass matrix $M'_N$ are negative in our analysis, and we regard the parities of these state as negative.

\subsection{One pion interactions and axial charges}

The interaction terms of nucleons to one pion are given from the Lagrangian as
\begin{equation}
\bar{N'}C'_{\pi NN}i\gamma_5\pi N'
+ \bar{N'}C'_{\partial \pi NN}\gamma^{\mu}\partial_{\mu}\pi \gamma_5N' \ , 
\end{equation}
where $\pi =\vec{\pi}\cdot\vec{\tau}$ and
\begin{equation}
C'_{\pi NN}=
\begin{pmatrix}
				-\frac{g_1}{2} &-\frac{y_1}{2\sqrt{6}} &0 & 0\\
				-\frac{y_1}{2\sqrt{6}} & \frac{5g_3}{12} & 0 & 0\\
				0 & 0 & -\frac{g_2}{2} & -\frac{y_2}{2\sqrt{6}}\\
				0 & 0 & -\frac{y_2}{2\sqrt{6}} & \frac{5g_4}{12}
\end{pmatrix} \ ,
\end{equation}
\begin{equation}
C'_{\partial \pi NN}=
\begin{pmatrix}
				0	&	0	&-\frac{a_1}{2f_\pi}	&-\frac{a_3}{2\sqrt{6}f_\pi}	\\
				0	&	0	&-\frac{a_4}{2\sqrt{6}f_\pi}	&\frac{5a_2}{12f_\pi}	\\
				-\frac{a_1}{2f_\pi}	&	-\frac{a_4}{2\sqrt{6}f_\pi}	&0	&0\\
				-\frac{a_3}{2\sqrt{6}f_\pi}	&	\frac{5a_2}{12f_\pi}&	0&	0\\
\end{pmatrix} \ .
\end{equation}
Axial-vector charge matrix is determined as
\begin{equation}
\bar{N'}G'_A\gamma^{\mu}\mathcal{A}_{\mu}\gamma_5N' \ ,
\end{equation}
where
\begin{equation}
G'_A=
\begin{pmatrix}
				-1	&	0	&-\frac{a_1\sigma_0}{f_\pi}	&-\frac{a_3\sigma_0}{\sqrt{6}f_\pi}	\\
				0	&	\frac{5}{3}	&-\frac{a_4\sigma_0}{\sqrt{6}f_\pi}	&\frac{5a_2\sigma_0}{6f_\pi}	\\
				-\frac{a_1\sigma_0}{f_\pi}	&	-\frac{a_4\sigma_0}{\sqrt{6}f_\pi}	&1	&0\\
				-\frac{a_3\sigma_0}{\sqrt{6}f_\pi}	&	\frac{5a_2\sigma_0}{6f_\pi}&	0&	-\frac{5}{3}\\
\end{pmatrix} \ .
\end{equation}
In the present analysis, we identify the mass eigenstates as
\begin{equation}
N^T_{\rm phys}\equiv(N(939)\,,\,N(1440)\,,\,N(1535)\hspace{5mm}N(1650))_{\rm phys}.
\end{equation}

\section{Chiral invariant masses and partner structure}
\label{sec:CIM}
In this section we determine the values of model parameters and study the mixing structure of relevant baryons.

As we said in the previous section, we set the VEV of $\sigma$ to be the pion decay constant:
\begin{equation}
\sigma_0=f_{\pi}=92.4{\rm MeV}
\end{equation}
Beside this parameter, there are twelve parameters in this model:
\begin{equation}
m_0^{(1)},m_0^{(2)},g_1,g_2,g_3,g_4,y_1,y_2,a_1,a_2,a_3,a_4
\end{equation}
We list values of relevant physical quantities determined from experiments and lattice analyses in Table.\ref{expvalue}.  Among them, we use the following  ten physical values as inputs: 
nucleon masses:
\begin{eqnarray}
&&m_{N(939)}=939{\rm MeV} \ , 
\nonumber
\\
&&m_{N(1440)}=1430{\rm MeV} \ ,
\nonumber
\\
&&m_{N(1535)}=1535{\rm MeV} \ ,
\nonumber
\\
&&m_{N(1650)}=1650{\rm MeV} \ ,
\label{mass values}
\end{eqnarray} 
partial decay widths:~\footnote{
In calculating the decay widths, we use the pion mass as 
$m_{\pi}=137{\rm MeV}$.
}
\begin{align}
&\Gamma(N(1440)\to N(939)+\pi)=228{\rm MeV}\ ,
\notag\\
&\Gamma(N(1535)\to N(939)+\pi)=68{\rm MeV}\ ,
\notag\\
&\Gamma(N(1650)\to N(939)+\pi)=84{\rm MeV}\ ,
\notag\\
&\Gamma(N(1650)\to N(1440)+\pi)=22{\rm MeV} \ ,
\label{width values}
\end{align} 
and axial charges: 
\begin{align}
&g_A(N(939))=1.272 \ ,
\notag\\
&g_A(N(1650)=0.55 \ .
\label{gA values}
\end{align}
\begin{table}
\begin{tabular}{|l|c|c|c|c|} \hline
& $P$ & Mass & Width[$\Gamma_{N^*\to N\pi}$] & axial charge
\\ \hline \hline
    N(939) & $+$ & 939.0$\pm$1.3 & - 			& 1.272$\pm$0.002 \\
    N(1440) & $+$ & 1430$\pm$20 & 228$\pm$74 & -\\
    N(1535) & $-$ & 1535$\pm$10 & 68$\pm$19 & $\mathcal{O}(0.1)$[lat]\\ 
    \multirow{2}{*}{N(1650)} & \multirow{2}{*}{$-$} & \multirow{2}{*}{1655$\pm$15} & 84$\pm23$[to N(939)] & \multirow{2}{*}{0.55[lat]}\\	
	&	&		&	22$\pm$15[to N(1440)]	&	\\	\hline
\end{tabular}
\caption{Experimental values of masses and partial decay widths of baryons listed in Ref.~\cite{PDGC}. 
The column indicated by $ P = \pm$ shows the parity of the nucleon.  Unit of masses and widths is MeV. 
The error of $m_{N(939)}$ expresses the mass difference 
between the proton and neutron. [lat] indicates that the value is obtained by the lattice analysis in Ref.~\cite{Takahashi:2008fy}. }
\label{expvalue}
\end{table}
In addition to the above inputs, we use the following range of $g_A(N(1535))$ to restrict the parameters:
\begin{equation}
-0.25\leq g_A(N(1535)) \leq 0.25.
\end{equation}
Furthermore, we restrict the parameters by requiring all the components of axial-charge matrix on the physical base are no larger than 5.

In this analysis, we first fix the chiral invariant masses $m_0^{(1)}$ and $m_0^{(2)}$ to certain values of 
every 5\,MeV from 0\,MeV to 1500\,MeV,  and determine other ten parameters from the values shown in  Eqs.~(\ref{mass values})-(\ref{gA values}). 
Here, $m_0^{(1)}$ and $m_0^{(2)}$ are chiral invariant masses  of $[(\bf{1},\bf{2})\oplus(\bf{2},\bf{1})]$  and $[(\bf{2},\bf{3})\oplus(\bf{3},\bf{2})]$ representations, respectively.

\begin{figure}
\begin{center}
\includegraphics[clip,width=8.0cm]{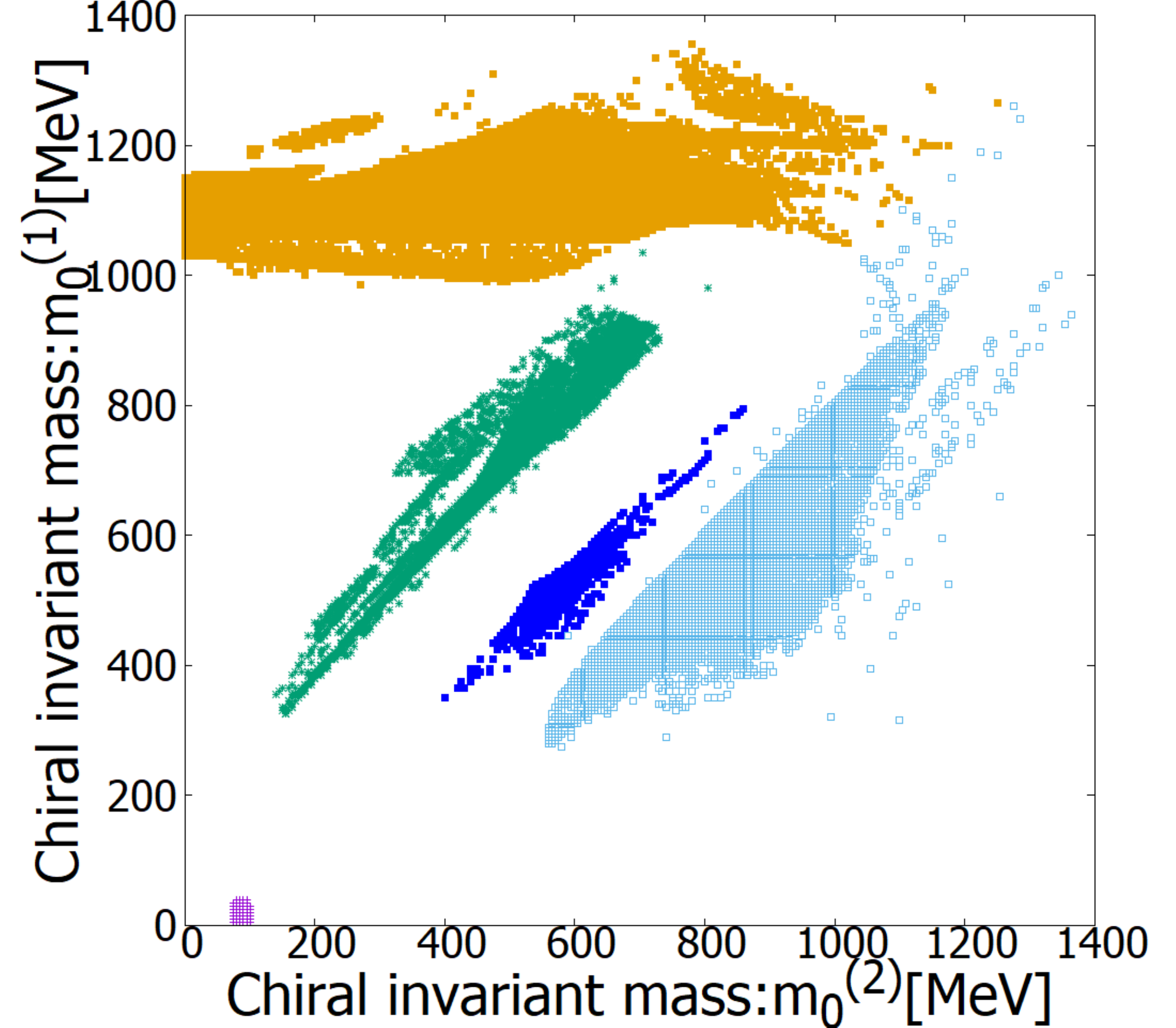}	
\end{center}
\caption{Chiral invariant masses for which we find solutions to reproduce the values in Eqs.~(\ref{mass values})-(\ref{gA values}). }
\label{chiralinvariantmass}
\end{figure}

We find  that solutions are categorized into five groups 
as shown
in Fig.\ref{chiralinvariantmass}.
In  Group 1  indicated by purple $+$ symbols, both  chiral invariant masses  are less than  $100$\,MeV.   
In the range  where $m_0^{(1)}$ is between $200$\,MeV and $900$\,MeV, three lumps exist: Group 2  indicated by blue $\blacksquare$ symbols;  Group 3  by light green $\times$ symbols;  and Group 4  light blue $\times\hspace{-7.5pt}+$ symbols. 
They are characterized by 
\begin{align}
m_0^{(1)}\sim m_0^{(2)} \ [{\rm Group 2}] \ , \notag\\
m_0^{(1)}\geq m_0^{(2)}\  [{\rm Group 3}] \ , \notag\\
m_0^{(1)}<m_0^{(2)} \ [{\rm Group 4}] \ ,	\notag\\
m_0^{(1)}\geq m_0^{(2)} \ [{\rm Group 5}] \ .
\end{align}
In Group 5 indicated by yellow $\square$, the chiral invariant mass $m_0^{(1)}$ takes large value of about 1000\,MeV. 

\begin{figure}[H]
\begin{center}
\includegraphics[clip,width=8.5cm]{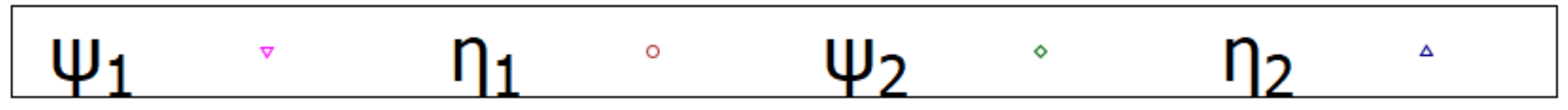} 
\\
\includegraphics[clip,width=8.9cm]{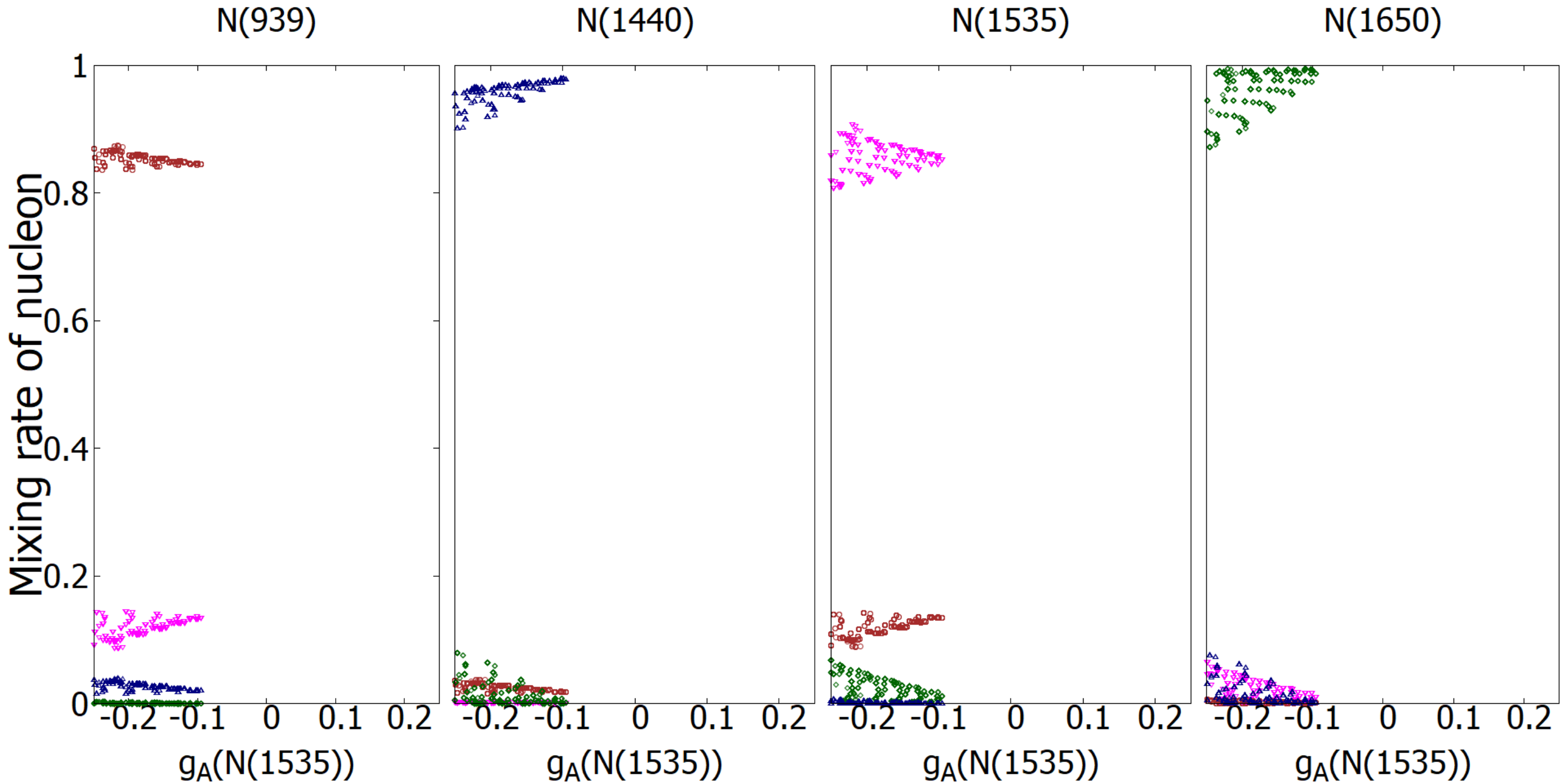} 
\\
\hspace{1mm}
\\	
\includegraphics[clip,width=8.9cm]{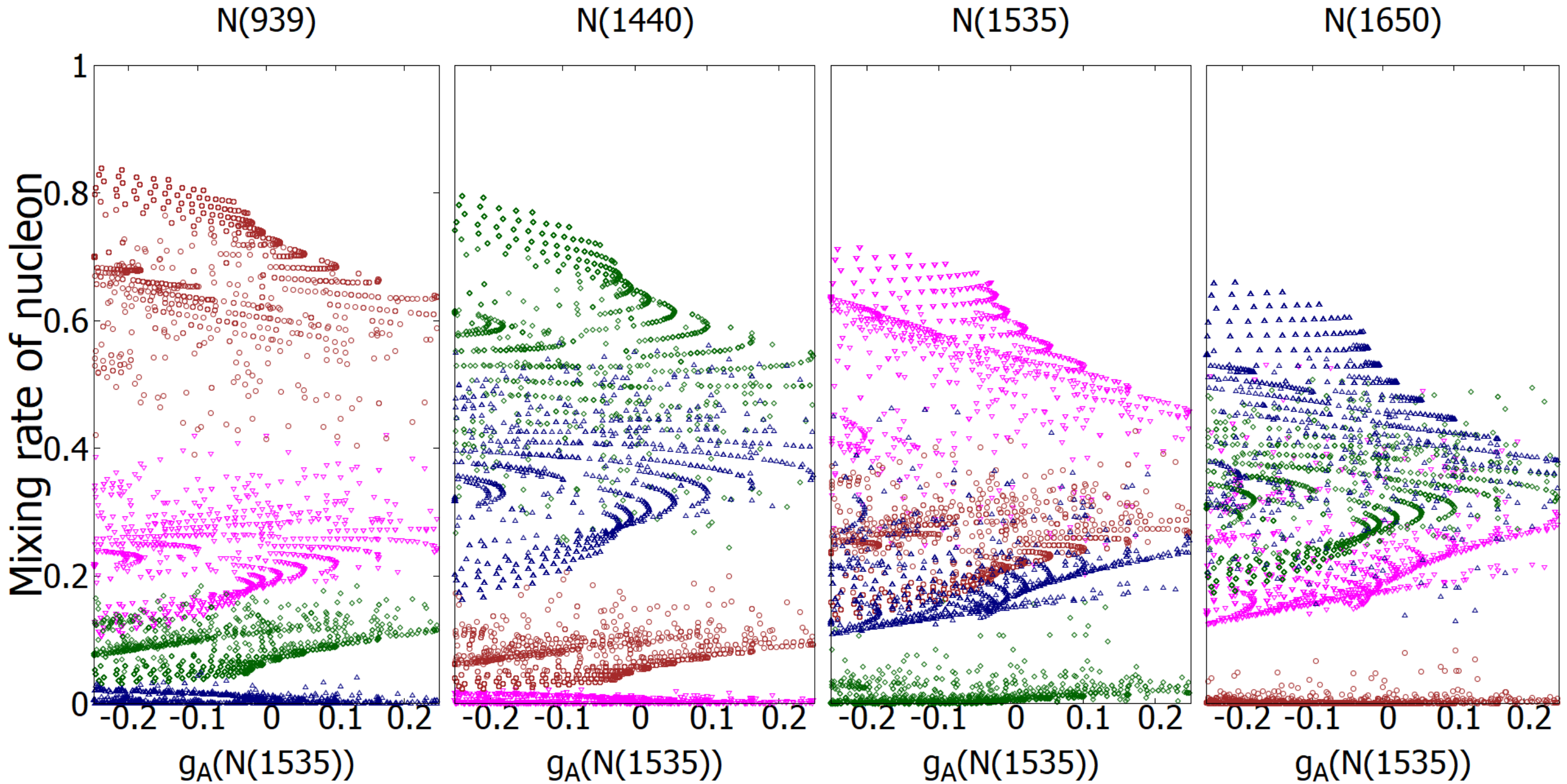} 
\\
\hspace{1mm}
\\
\includegraphics[clip,width=8.9cm]{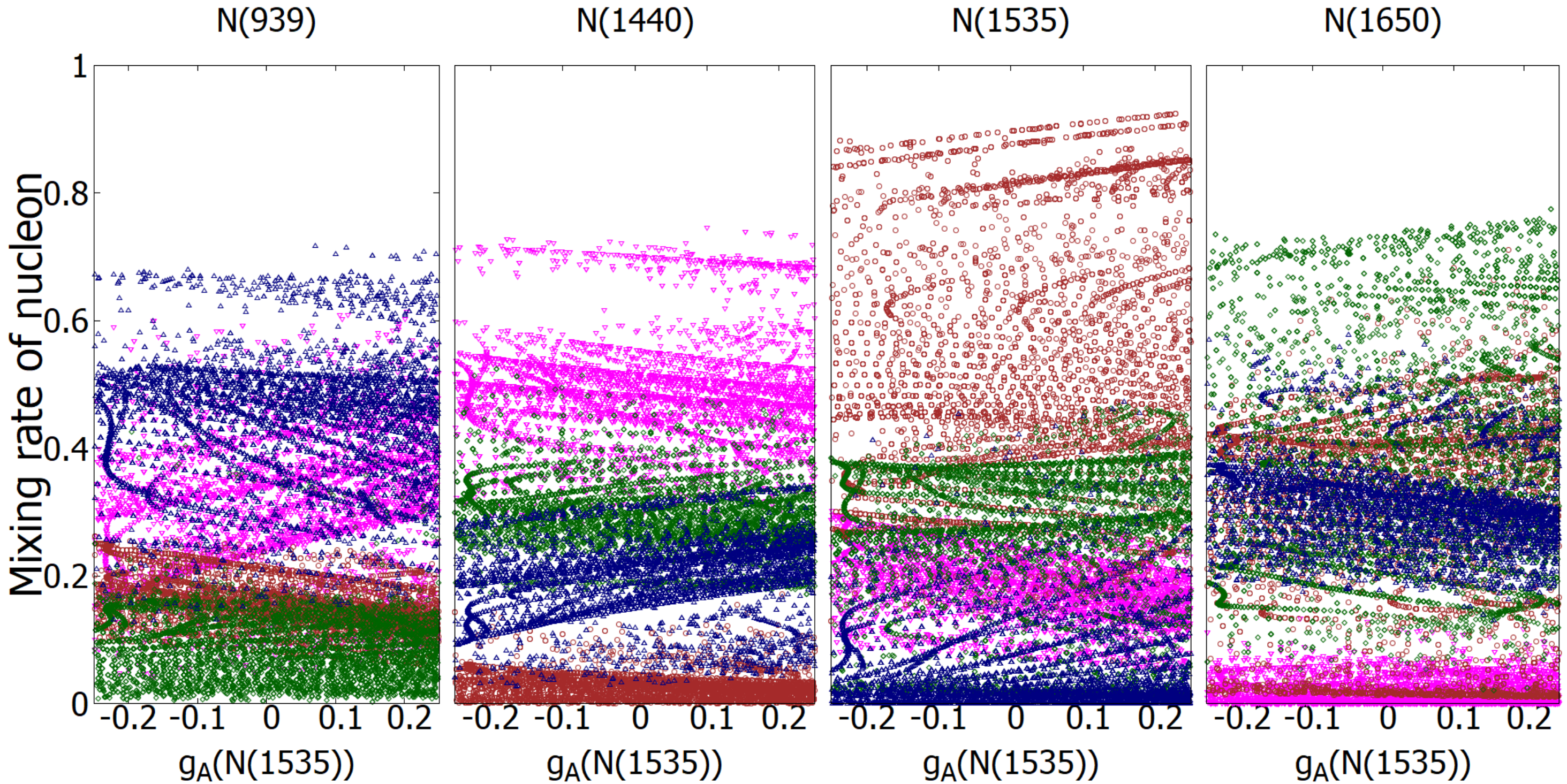}
\\
\hspace{1mm}
\\
\includegraphics[clip,width=8.9cm]{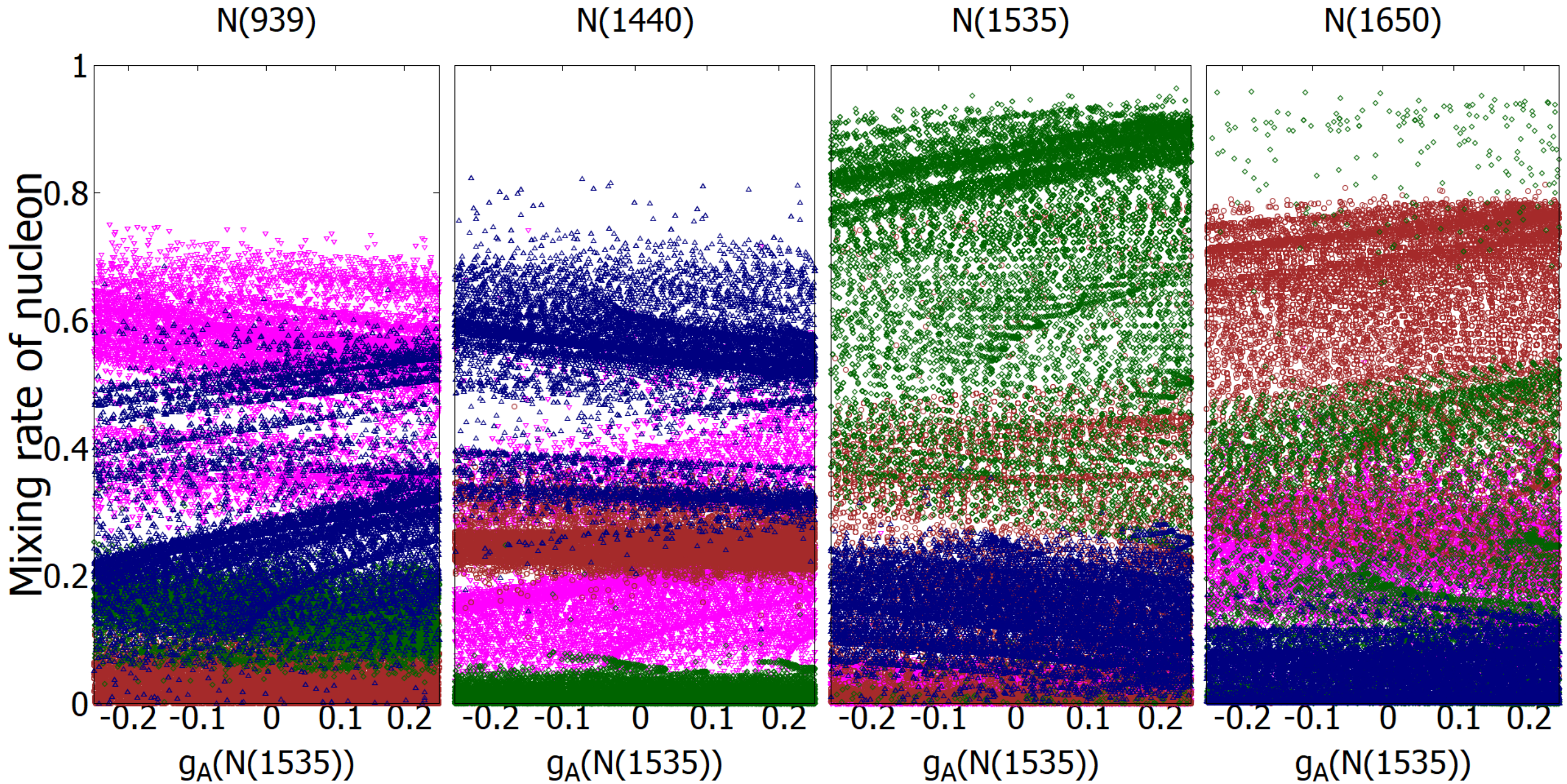} 
\\
\hspace{1mm}
\\	
\includegraphics[clip,width=8.9cm]{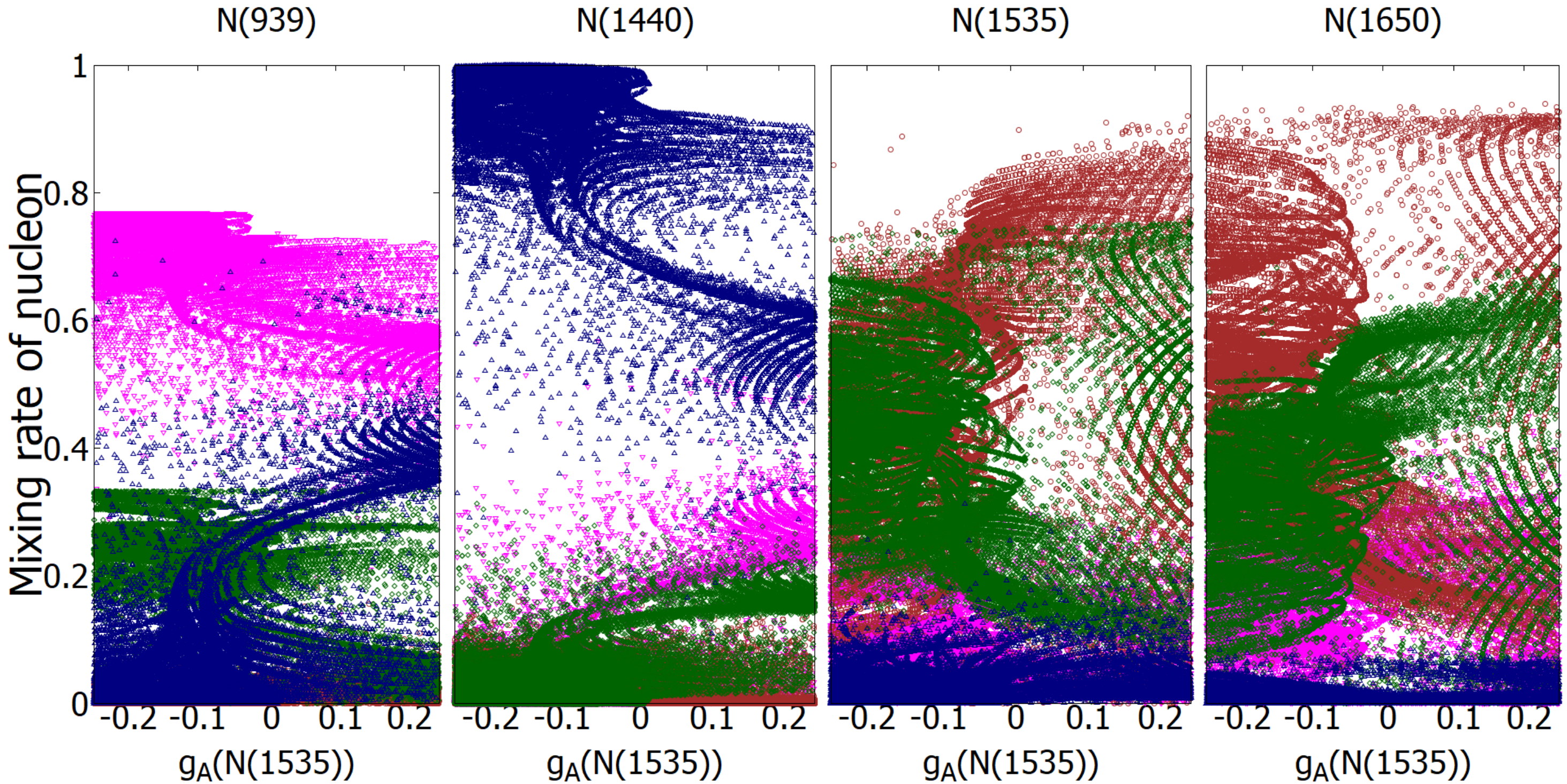} 
\end{center}
\caption{Mixing rates of nucleons. The first to fifth rows shows Group 1 to Group 5.}
\label{mixingrate}
\end{figure}

In Fig.~\ref{mixingrate}, we show the mixing structure of nucleons: $N(939)$, $N(1440)$, $N(1535)$ and $N(1650)$ for Group 1 to Group 5.  Here the horizontal axis shows the value of  axial-charge of $N(1535)$  and 
the vertical axis shows the percentages of $\psi_1$ indicated by magenta $\triangledown$ symbols, $\eta_1$ by brown $\circ$ symbols, $\psi_2$ by green $\diamond$ symbols and 
$\eta_2$ by navy $\vartriangle$ symbols.
In Table~\ref{nucleonstate}, we summarize features of mixing rates for each group. 
The first row in  Fig.~\ref{mixingrate}  shows that 
the dominant component of 
$N(939)$ is  $\eta_1$  indicated by brown $\circ$  belonging to $[(\bf{2},\bf{3})\oplus(\bf{3},\bf{2})]$ representation. 
We note that we cannot find any solutions for $g_A(N(1535))\gtrsim -0.1$ in the  Group 1. 
One can easily see that $N(1440)$ is dominated by  $\eta_2$  (navy $\vartriangle$),
$N(1535)$ by  $\psi_1$(magenta $\triangledown$)  belonging to  $[(\bf{1},\bf{2})\oplus(\bf{2},\bf{1})]$ representation and $N(1650)$ bv  $\eta_2$ (green $\diamond$). 
Since $\eta_1$ and $\eta_2$ are chiral partners to each other, we conclude that $N(1440)$ dominated by $\eta_2$ is the chiral partner to $N(939)$ dominated by $\eta_1$.
We would like to stress that this partner structure can be realized when the chiral invariant masses of $N(939)$ and $N(1440)$ are small. 

In Group 2 (the second row of Fig.~\ref{mixingrate}), 
$\eta_1$ (brown $\circ$) belonging to $[(\bf{2},\bf{3})\oplus(\bf{3},\bf{2})]$ representation is a dominant component in the $N(939)$
and $\psi_1$ (magenta $\triangledown$) almost occupies $N(1535)$,  similarly to Group 1.
A  difference between Group 1 and Group 2  appears in the rate of $\psi_2$  (green $\diamond$)  in $N(1440)$. In Group 1, the mixing rate 
of $\psi_2$ in $N(1440)$ is smaller than $0.1$ as can be seen  in the first  row of Fig.~\ref{mixingrate}. On the other hand,  the rate of $\psi_2$ is larger than $0.2$ and $\psi_2$ is included in $N(1440)$ dominantly  as shown  in the second  row of  Fig.~\ref{mixingrate}. 
Here the rate of $\eta_2$ component (navy $\vartriangle$)  included in $N(1650)$ is high, but $N(1440)$ and $N(1535)$ include a certain amount of the $\eta_2$ component.
So, it is difficult to identify the chiral partner of $N(939)$ in Group 2.
\begin{table}
\begin{tabular}{|c|c|c|c|c|c|} \hline
Group	&	Group 1	&	Group 2	&	Group 3	&	Group 4	&	Group 5
\\
color	&	purple	&	blue		&	light green	&	sky		&	yellow		
\\ \hline \hline
$N(939)$		&	$\eta_1>0.8$	&	$\eta_1>0.35$	&	$\eta_1<0.45$	&	$\eta_1<0.35$	&	$\eta_1<0.35$
\\ \hline
\multirow{3}{*}{$N(1440)$}	&	$\psi_1<0.01$	&	$\psi_1<0.025$	&	$\psi_1>0.25$		&	$\psi_1<0.75$	&	$\psi_1<0.6$
\\
		&	$\psi_2<0.1$	&	$\psi_2>0.2$	&	$\psi_2>0.15$		&	$\psi_2<0.2$		&	$\psi_2<0.35$
\\ \hline
$N(1650)$	&	$\eta_2<0.1$	&	$\eta_2>0.1$	&	$\eta_2>0.1$	&	$\eta_2<0.3$	&	$\eta_2\leq0.1$
\\ \hline
\end{tabular}
\caption{Features of mixing rates. For example, in the column for Group 1, $\eta_1 > 0.8$ in the row of $N(939)$ implies that the percentage of $\eta_1$ component in $N(939)$ is always larger than $0.8$.}
\label{nucleonstate}
\end{table}

In Group 3, Group 4 and Group 5, $N(939)$ is composed of $\psi_1$(magenta $\triangledown$) or $\eta_2$(navy $\vartriangle$) dominantly and negative parity nucleons,
$N(1535)$ and $N(1650)$ have  $\psi_2$ (green $\diamond$)) or $\eta_2$ (brown $\circ$) mainly,  as can be seen in the third, fourth and fifth rows in Fig.~\ref{mixingrate}. 
This indicates that the chiral partner of $N(939)$ is  a mixture of two negative parity nucleons in these groups, differently from Group 1 and Group 2. 
Table~\ref{nucleonstate} shows that Group 5 is distinguished from Group 3 by the mixing rate of $\eta_2$ 
component in $N(1650)$:
The rate is  larager  than $0.1$  in Group 3, while it is no greater than $0.1$ in Group 5. 
On the other hand, it is difficult to distinguish Group 4 with Group 5 and Group 3 with Group 4 by mixing rates.
In the present work, we use the values of chiral invariant masses in addition to the mixing rates to separate these Groups: The chiral invariant masses satisfy $m_0^{(1)}<m_0^{(2)}$ in Group 4;  Group 3 is characterized by  $m_0^{(1)}\geq m_0^{(2)}$ and $\eta_2>0.1$ in $N(1650)$,  while Group 5 by $m_0^{(1)}\geq m_0^{(2)}$ and $\eta_2\leq0.1$ in $N(1650)$. 
In Ref.~\cite{Olbrich:2015gln}, two $[(\bf{1},\bf{2})\oplus(\bf{2},\bf{1})]$ representations are used to study $N(939)$, $N(1440)$, $N(1535)$ and $N(1650)$.
The authors found two types of  solutions.  In one type (Type-A), One is that  the chiral partner of $N(939)$ is a mixture of $N(1535)$ and $N(1650)$ in vacuum and both chiral invariant masses are almost $200$\,MeV. 
In another type (Type-B) $N(939)$ and  $N(1535)$ are chiral partners to each other and  both chiral invariant masses are about $1000$\,MeV. 
In the present analysis, we find solutions for which $N(939)$ is dominated by $[(\bf{1},\bf{2})\oplus(\bf{2},\bf{1})]$ representation in Groups 3, 4 and 5.
Comparing mixing structure and the chiral invariant masses, we think that the Type-A is consistent with Group 3 and Type-B with Group 4 or 5. 

In Ref.~\cite{Nishihara:2015fka}, $[(\bf{3},\bf{\bar{3}})\oplus(\bf{\bar{3}},\bf{3})]$ and $[(\bf{3},\bf{6})\oplus(\bf{6},\bf{3})]$ representations under the chiral $U(3)_{\rm L}\otimes U(3)_{\rm R}$ symmetry are introduced  to study  six nucleons including $N(939)$, $N(1440)$, $N(1535)$ and $N(1650)$. This analysis indicates that $N(939)$ have $[(\bf{3},\bf{6})\oplus(\bf{6},\bf{3})]$ dominantly and the chiral invariant mass is 500MeV$\sim$800MeV. This is consistent with Group 2 in our analysis.

\begin{table}[htb]
\begin{tabular}{|c|c|c|c|c|c|c|} 
\hline
Nucleon				&	rep		&	Group1	&	Group 2		&	Group 3	&	Group 4	&	Group 5		
\\ \hline
\multirow{4}{*}{$N(939)$}&	$\psi_1$	&	0.113		&	0.21			&	\cellcolor{yellow}0.518		&	\cellcolor{yellow}0.677		&\cellcolor{yellow}0.768
\\
					&	$\eta_1$	&	\cellcolor{yellow}0.856		&	\cellcolor{yellow}0.735		&	0.098		&	0.032		&0.001
\\
					&	$\psi_2$	&	0.002		&	0.052		&	0.133		&	0.163		&0.227
\\
					&	$\eta_2$	&	0.029		&	0.002		&	0.251		&	0.128		&0.004	
\\ \hline
\multirow	{4}{*}{$N(1440)$}&	$\psi_1$	&	0.002		&	0.003		&	\cellcolor{yellow}0.355		&	0.092		&0.002	
\\
					&	$\eta_1$	&	0.029		&	0.043		&	0.021		&	0.223		&0.041
\\
					&	$\psi_2$	&	0.033		&	\cellcolor{yellow}0.627		&	\cellcolor{yellow}0.368		&	0.01			&0.001
\\
					&	$\eta_2$	&	\cellcolor{yellow}0.936		&	0.327		&	0.256		&	\cellcolor{yellow}0.675		&\cellcolor{yellow}0.956	
\\ \hline
\multirow{4}{*}{$N(1535)$}&	$\psi_1$	&	\cellcolor{yellow}0.819		&	\cellcolor{yellow}0.641		&	0.126		&	0.012		&0.137
\\
					&	$\eta_1$	&	0.109		&	0.222		&	\cellcolor{yellow}0.801		&	0.212		&0.322
\\
					&	$\psi_2$	&	0.068		&	0.007		&	0.057		&	\cellcolor{yellow}0.615		&\cellcolor{yellow}0.528
\\
					&	$\eta_2$	&	0.004		&	0.130		&	0.015		&	0.161		&0.013	
\\ \hline
\multirow	{4}{*}{$N(1650)$}&	$\psi_1$	&	0.066		&	0.146		&	0.001		&	0.218		&0.093
\\
					&	$\eta_1$	&	0.006		&	0.0			&	0.08			&	\cellcolor{yellow}0.533		&\cellcolor{yellow}0.636
\\
					&	$\psi_2$	&	\cellcolor{yellow}0.897		&	0.314		&	\cellcolor{yellow}0.442		&	0.212		&0.244
\\
					&	$\eta_2$	&	0.031		&	\cellcolor{yellow}0.540		&	\cellcolor{yellow}0.477		&	0.037		&0.027	
\\ \hline
\end{tabular}
\caption{Typical values of mixing rates of nucleons. }
\label{tablemixingrate}
\end{table}

\section{Axial charges}
\label{sec:axial charge}

In  the previous  section, we discussed the mixing structure of nucleons together with their chiral invariant masses. 
In this section, we study axial charges in detail.

We define transition axial charge as off-diagonal elements of following axial-charge matrix on physical base.
\begin{equation}
\bar{N'}_{\rm phys}G'_{A {\rm phys}}\gamma^{\mu}\mathcal{A}_{\mu}\gamma_5N'_{\rm phys}
\end{equation}
where
\begin{equation}
G'_{A {\rm phys}}=
\begin{pmatrix}
				g_A(N_1)		&	g_A(N_1N_2)	&	g_A(N_1N_3)	&g_A(N_1N_4)	\\
				g_A(N_1N_2)	&	g_A(N_2)	&	g_A(N_2N_3)	&g_A(N_2N_4)	\\
				g_A(N_1N_3)	&	g_A(N_2N_3)	&     g_A(N_3)	&g_A(N_3N_4)\\
				g_A(N_1N_4)	&	g_A(N_2N_4)	&	g_A(N_3N_4)	&g_A(N_4)\\
\end{pmatrix}
\end{equation}
and
\begin{equation}
(N_1,N_2,N_3,N_4)\equiv(N'^{(1)}_+\hspace{5mm}N'^{(2)}_+\hspace{5mm}N'^{(1)}_-\hspace{5mm}N'^{(2)}_-)_{\rm phys}
\end{equation}
\begin{figure}[H]
\begin{center}
\includegraphics[clip,width=9.0cm]{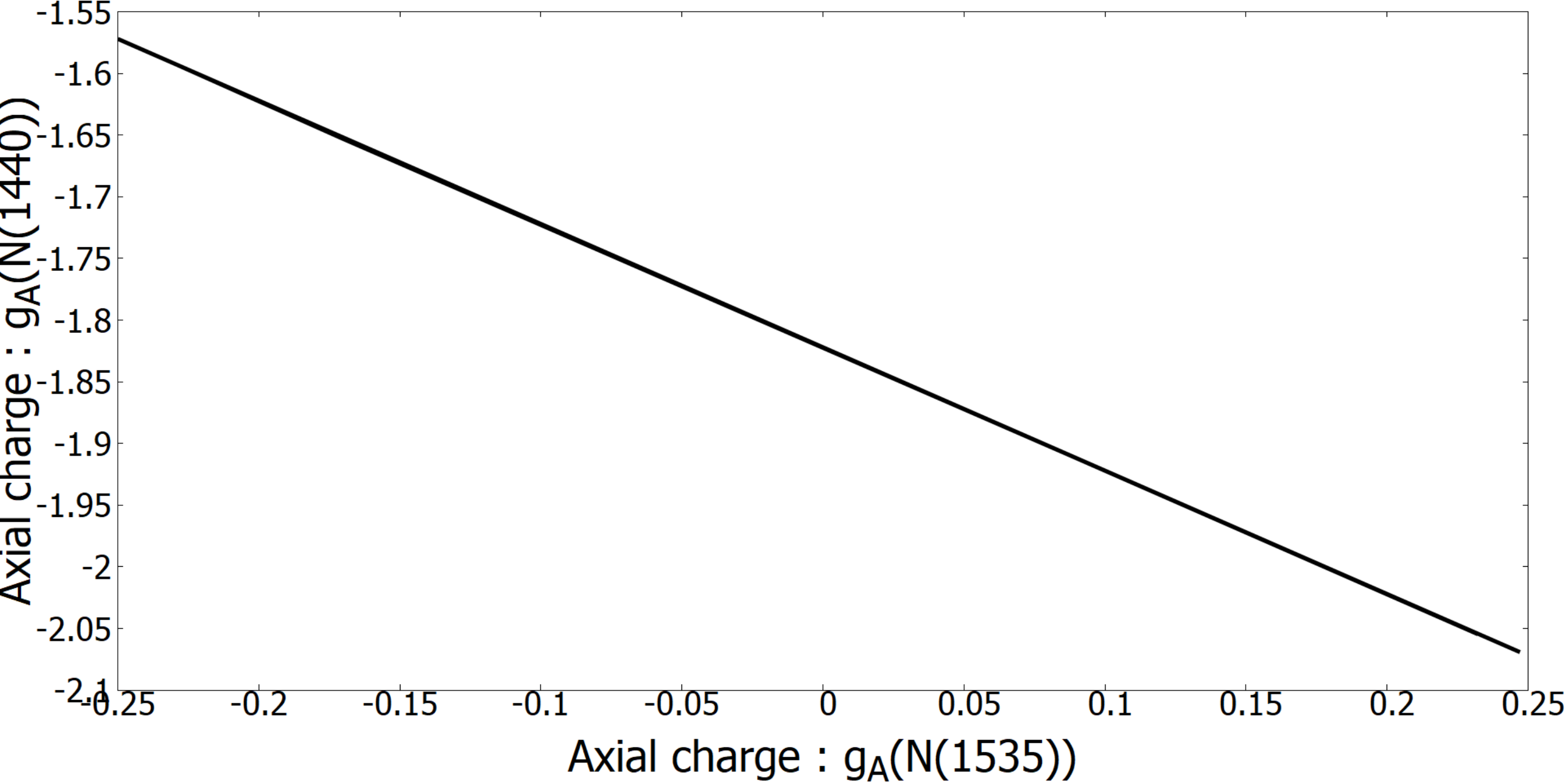}	
\end{center}
\caption{The values of $g_A(N_2)$ and $g_A(N_3)$:points of all Groups. This relation is determined by $\sum_{i=1}^4g_A(N_i)=0$.}
\label{gAN1440}
\end{figure}
In the present model, the following relation the diagonal axial charges is satiefied: 
\begin{equation}
\sum_{i=1}^4g_A(N_i)=0
\end{equation}
Now we use axial charge:$g_A(N_1)=g_A(N(939))=1.272$ and $g_A(N_4)=g_A(N(1650))=0.55$ as input. So this relation is
\begin{equation}
g_A(N_2)+g_A(N_3)=-1.822 .
\end{equation}
We plot this relation in Fig.\ref{gAN1440}. 
This plot shows that, when the axial charge of $N1535)$ is in the range consistent with the lattice analysis, the axial charge of $N(1440)$ is negative.
\begin{figure}[H]
\begin{center}
\includegraphics[clip,width=9.0cm]{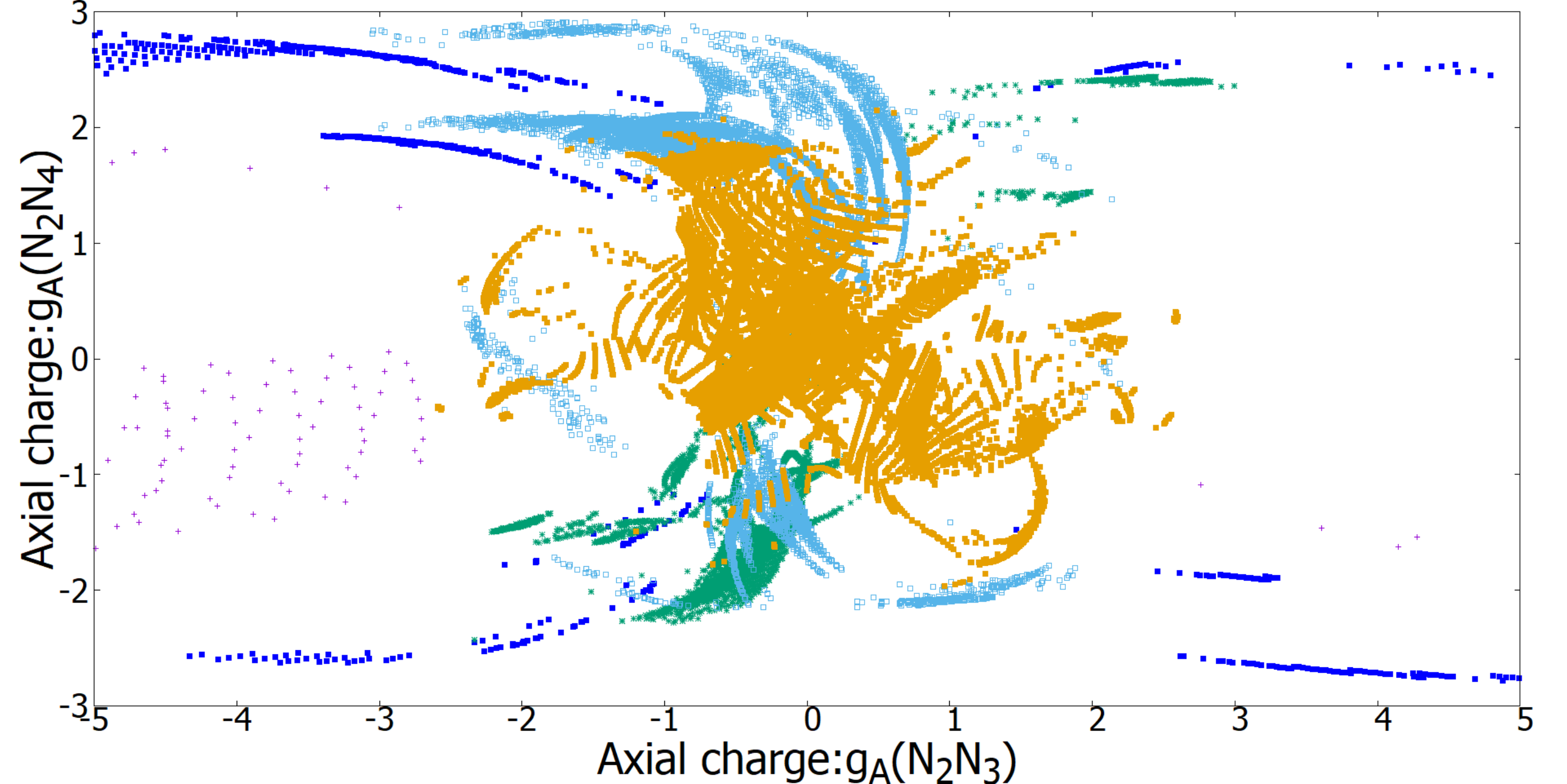}	
\end{center}
\caption{Predicted values of $g_A(N_2N_3)$ and $g_A(N_2N_4)$.}
\label{gAN2N3-gAN2N4}
\end{figure}
In Fig.~\ref{gAN2N3-gAN2N4} we plot predicted values of $g_A(N_2N_3)$ and $g_A(N_2N_4)$. This shows 
that $|g_A(N_2N_3)|$ of Group 1 and Group 2  is always larger than $2$ and $0.5$, respectively. 
On the other hand, $|g_A(N_2N_4)|$ of Group 1 is smaller than 2 and that of Group 2 is above 1. We see that Group 1 is able to be distinguished from other groups.  
\begin{figure}[H]
\begin{center}
\includegraphics[clip,width=9.0cm]{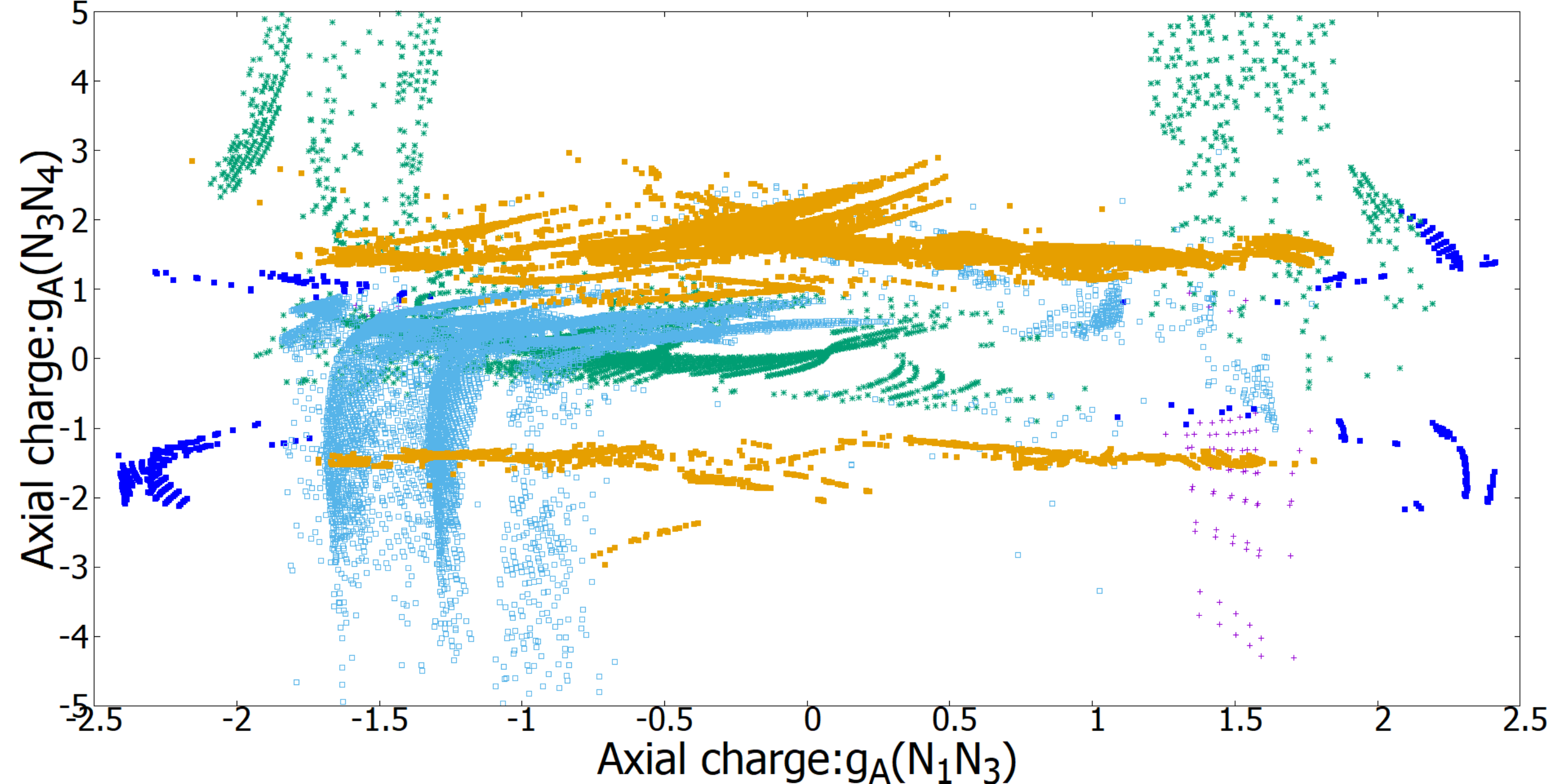}	
\end{center}
\caption{Predicted values of $g_A(N_3N_4)$ and $g_A(N_1N_3)$. }
\label{gAN3N4-gAN1N3}
\end{figure}
Predicted values of $g_A(N_3N_4)$ and $g_A(N_1N_3)$ are plotted in Fig.~\ref{gAN3N4-gAN1N3}. 
We note that $|g_A(N_1N_3)|$ belonging to Group 1 and Group 2 are larger than $1$, and  that $|g_A(N_3N_4)|$ of Group 2  and Group 5 lies  between  $1$  and  $3$ and that of Group 1 is above $1$. In particular,  $|g_A(N_3N_4)|\sim2$ in Group 5.
\begin{figure}[H]
\begin{center}
\includegraphics[clip,width=9.0cm]{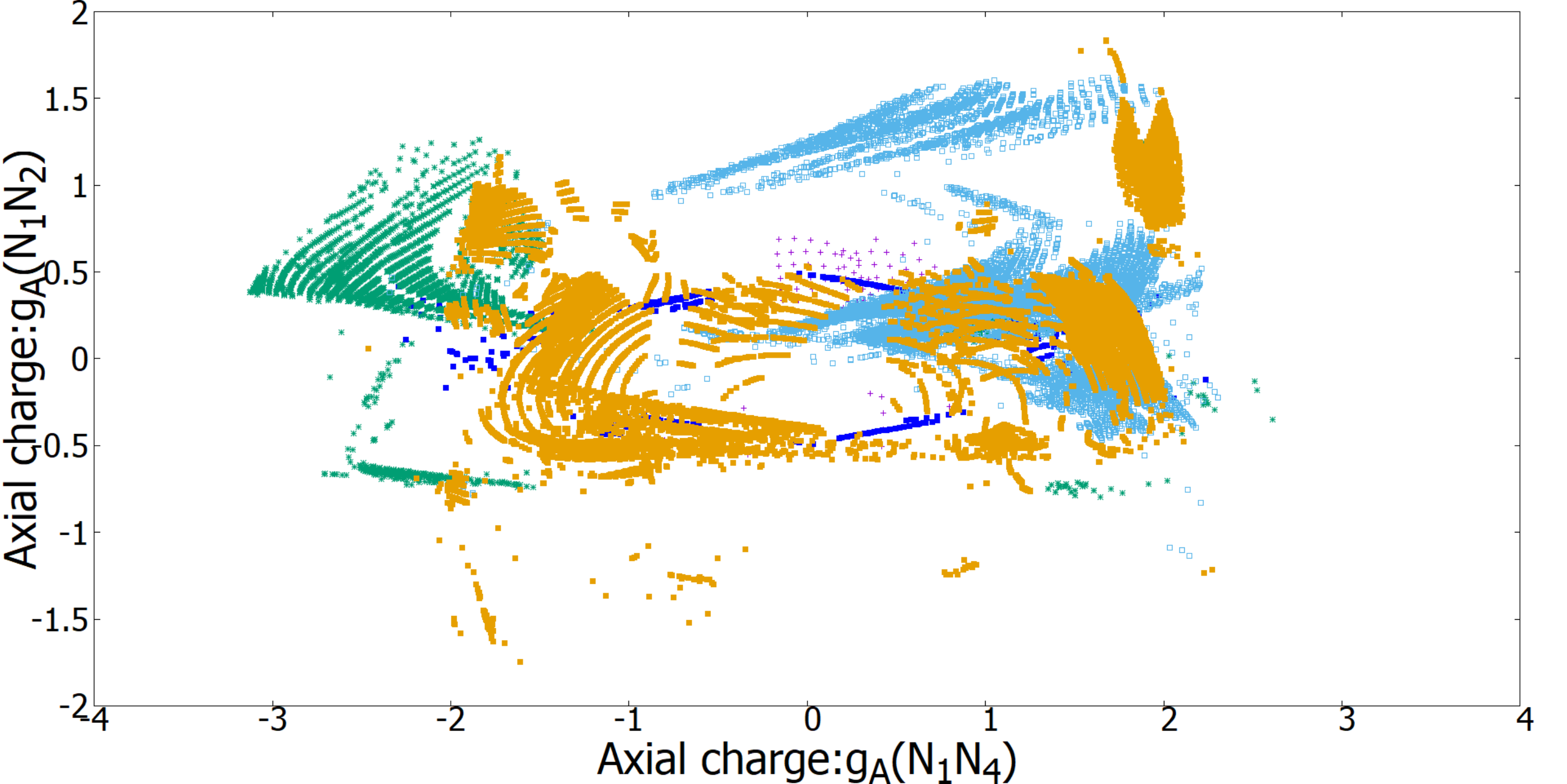}	
\end{center}
\caption{Predicted values of $g_A(N_1N_2)$ and $g_A(N_1N_4)$: There are no points of Group 3 in the vicinity of  $|g_A(N_1N_4)|\sim 0$. Values of Group 1, 2, 4 and 5 are mixed there. }
\label{gAN1N2-gAN1N4}
\end{figure}
We plot the values of $g_A(N_1N_2)$ and $g_A(N_1N_4)$ in Fig.~\ref{gAN1N2-gAN1N4}. From Fig.\ref{gAN1N2-gAN1N4}, we can find that $|g_A(N_1N_2)|$ belonging to Group 1 (purple $+$ symbols) and Group 2  (blue $\blacksquare$ symbols) are no larger than $1$, while that  belonging to Group 3 (light green $\times$ symbols) and Group 4 light blue $\times\hspace{-7.5pt}+$ symbols
are below $1.5$. $|g_A(N_1N_4)|$ belonging to Group 3 (light green $\times$ symbols) are larger than $0.5$.   

We summarize  typical predicted values of transition axial-charges and the range of transition axial-charges in Table.~\ref{tableaxialcharge}. 
\begin{table}
\scalebox{0.65}[1.0]
{
\begin{tabular}{|c|c|c|c|c|c|c|c|} \hline 
group	&	($m_0^{(1)}$ , $m_0^{(2)}$)	& $g_A(N_1N_2)$	&	$g_A(N_1N_3)$	&	$g_A(N_1N_4)$	&	$g_A(N_2N_3)$	&	$g_A(N_2N_4)$	&	$g_A(N_3N_4)$		
\\ \hline
1	&	(5 , 100)	&	0.689	&	1.707	&	-0.159	&	-4.512	&	-0.197	&	-4.303
\\ 
1	&	(40 , 95)	&	-0.505	&	-1.353	&	0.417	&	-2.859	&	1.310	&	0.804
\\
:1:	&	- 		&	$|g_A | < 1$	&	 $1<|g_A |<2$		&	$|g_A | < 1$	&	$|g_A |>2.5$	&	$|g_A |<2$	& $|g_A|>0.5$
\\ \hline
2	&	(510 , 540)	&	0.449	&	-2.224	&	0.247	&	-4.644	&	2.614	&	-1.979
\\
2	&	(645 , 715)&	-0.434	&	1.868	&	-1.132	&	-1.039	&	2.204	&	-0.921
\\
2	&	(790 , 855)&	0.252	&	-1.912	&	-1.306	&	-1.294	&	-1.614	&	1.238
\\
:2:	&	- 		&	$|g_A|<1$	&	$|g_A|>1$		&	$|g_A|<2.5$	&	$|g_A|>0.5$	&$|g_A|>1$	&	$|g_A|>0.5$
\\ \hline
3 	 &	(400 , 175)	&	0.906	&	-1.197	&	-1.697	&	0.935	&	-1.394	&	1.561
\\
3 	 &	(695 , 330)	&	0.206	&	1.549	&	0.888	&	2.825	&	2.394	&	4.95
\\
3 	 &	(940 , 620)	&	0.203	&	1.963	&	1.597	&	1.197	&	1.324	&	-0.242
\\
:3:	&	- 			&$|g_A|<1.5$	&	$|g_A|<2.5$		&	$|g_A|<4$	&	$|g_A|<3$	&$|g_A|<2.5$	&	$g_A>-1$
\\ \hline
4	 &	(365 , 685)	&	1.328	&	-1.694	&	0.175	&	-2.219	&	2.028	&	-1.419
\\ 
4	&	(650 , 945)	&	-0.099	&	-0.149	&	1.959	&	-0.534	&	-1.386	&	0.795
\\ 
4	&	(1070 , 1150)	&	-0.755	&	-0.303	&	2.136	&	-0.293	&	-0.159	&	1.714
\\
:4:	&	- 			&$|g_A|<1.5$	&	$|g_A|<2$		&$g_A<2.5$		&	$|g_A|<3.5$	&	$|g_A|<3$	&	$g_A<2$
\\ \hline
5	&	(1070 , 50)	&	-0.514	&	-1.350	&	-1.454	&	0.075	&	-0.548	&	-1.415
\\
5	&	(1085 , 500)&	0.713	&	1.210	&	-1.599	&	-0.058	&	0.107	&	1.492
\\ 
5 	&	(1210 , 1050)&	-0.295	&	0.175	&	1.788	&	-0.245	&	0.116	&	1.867
\\
:5:	&	- 			&$|g_A|<2$		&	$|g_A|<2.5$		&$|g_A|<2.5$	&	$|g_A|<3$	&	$|g_A|<2$	&$1<|g_A|<3$
\\ \hline
\end{tabular}
}
\caption{
Typical predicted values of axial charges. Rows indicated by :$j$: ($j=1,\ldots,5$) show that the absolute value of axial charge lie in the range shown in the rows.}
\label{tableaxialcharge}
\end{table}
\section{change of nucleon masses}
\label{sec:masses}

\begin{figure}[H]
\begin{center}
\includegraphics[clip,width=9.0cm]{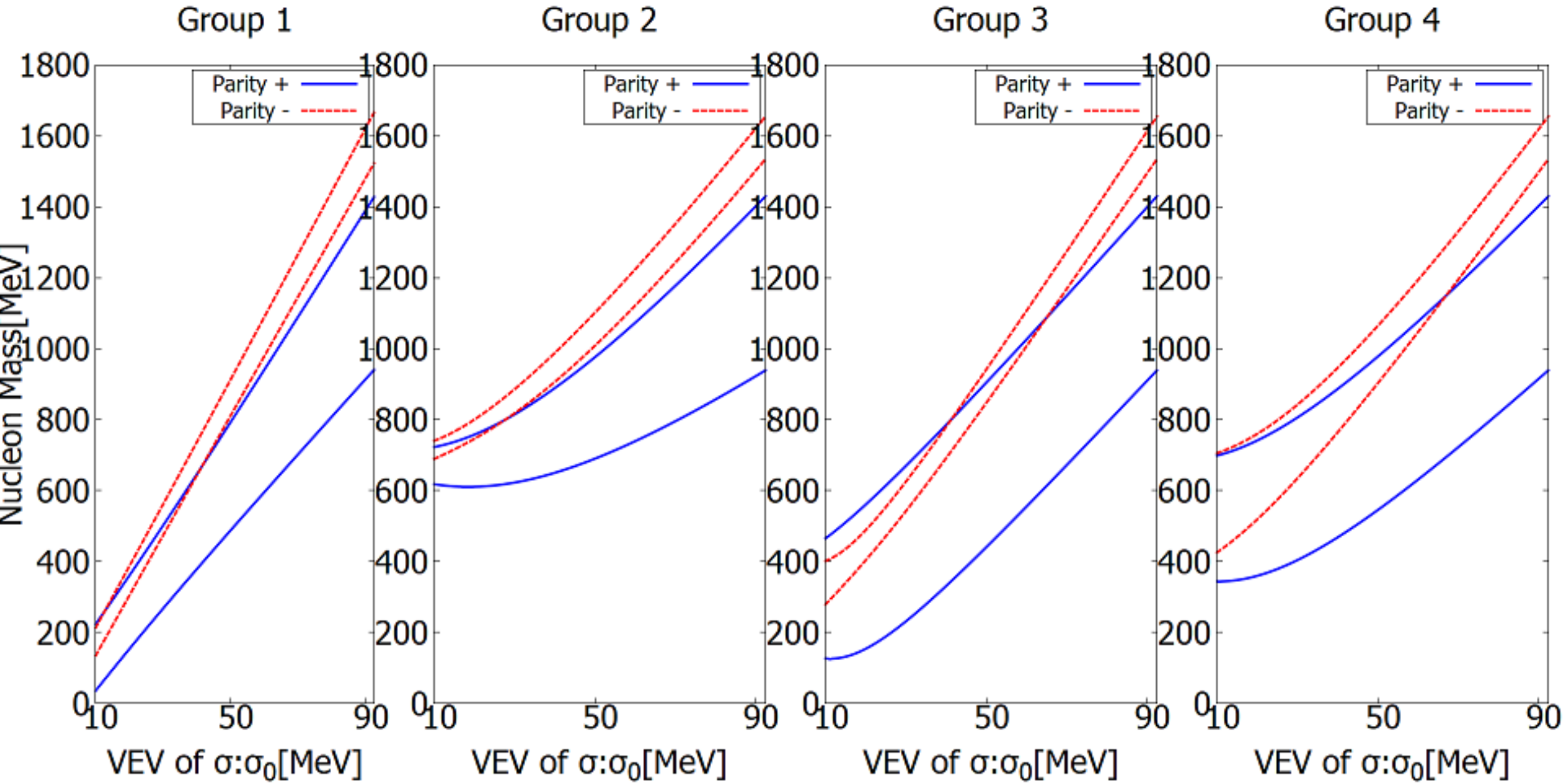}	
\end{center}
\caption{
Dependences of nucleon masses on the VEV of $\sigma$ for Groups 1-4.Values of chiral invariant masses used are 
 $(m_0^{(1)},m_0^{(2)})=(40{\rm MeV},95{\rm MeV})$ (Group 1), 
 $(m_0^{(1)},m_0^{(2)})=(645{\rm MeV},715{\rm MeV})$ (Group 2), 
$(m_0^{(1)},m_0^{(2)})=(400{\rm MeV},175{\rm MeV})$ (Group 3) and 
$(m_0^{(1)},m_0^{(2)})=(365{\rm MeV},685{\rm MeV})$ (Group 4). }
\label{mass_1}
\end{figure}

\begin{figure}[H]
\begin{center}
\includegraphics[clip,width=9.0cm]{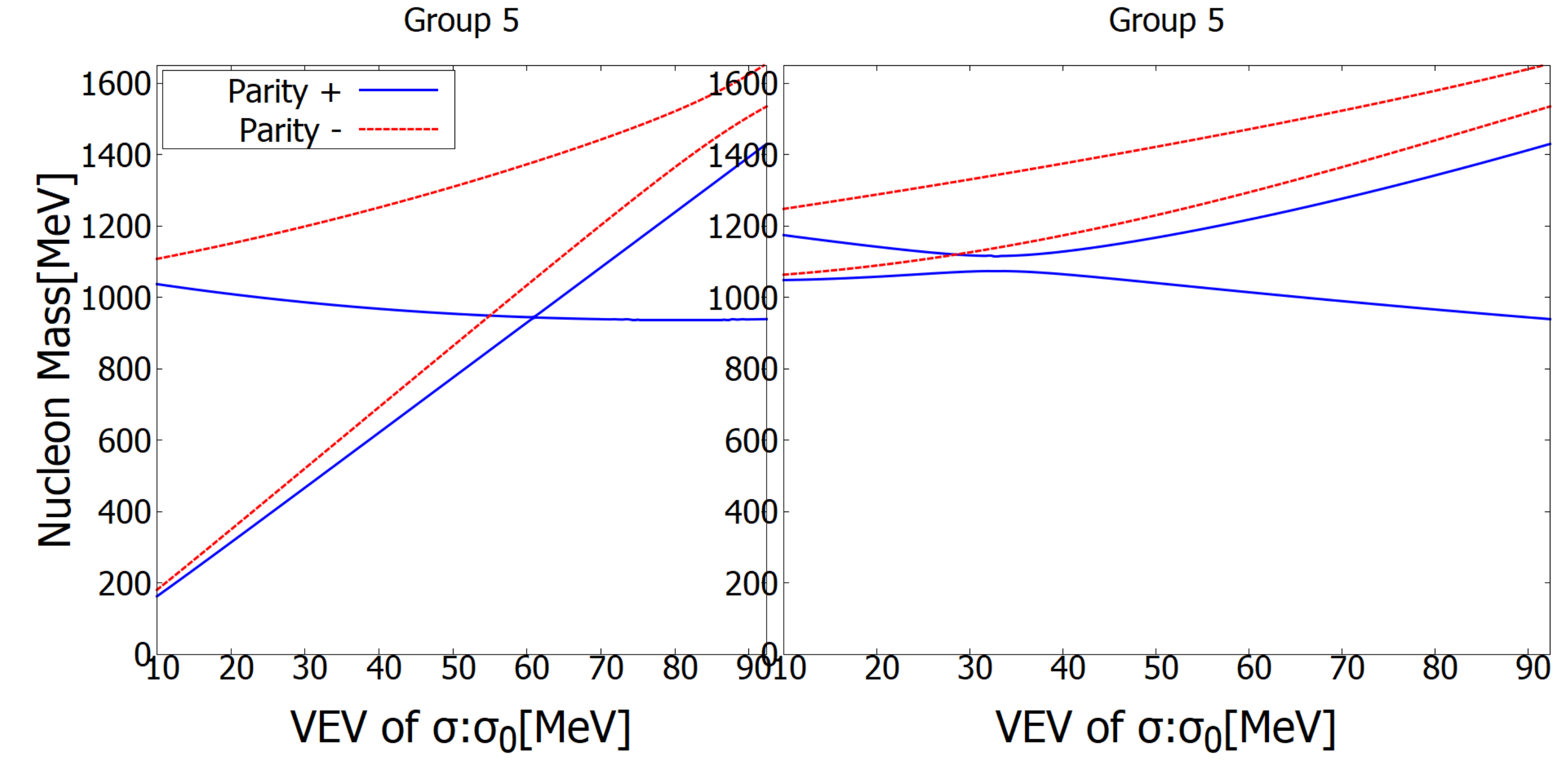}	
\end{center}
\caption{Dependences of nucleon masses on the VEV of $\sigma$ in Group 5 for 
$(m_0^{(1)}, m_0^{(2)})=(1070\mbox{MeV},50\mbox{MeV})$ (left figure) and  $(1210\mbox{MeV},1050\mbox{MeV})$ (right figure).
}
\label{mass_2}
\end{figure}

In this section we study the change of nucleon masses  when the VEV of $\sigma$ changed .
We plot the dependences of nucleon masses on the value of the VEV for Groups 1-4 in Fig.~\ref{mass_1} and those for Group 5 in Fig.~\ref{mass_2} for some choices of the chiral invariant masses, $m_0^{(1)}$ and $m_0^{(2)}$.
Figure~\ref{mass_1} shows that nucleons masses are decreased as $\sigma_0$ is decreased in Group 1, 2, 3 and 4,
if chiral invariant masses are smaller than mass of $N(939)$. 
We note that, for some parameter choices in Group 4, mass of ground state nucleon is increased as $\sigma_0$ is decreased.

In the case of Group 5 shown in Fig.~\ref{mass_2}, the value of $m_0^{(1)}$ is about $1000$\,MeV, while $m_0^{(2)}$ takes values in wide range.
Left panel of FIg.~\ref{mass_2} [$(m_0^{(1)}, m_0^{(2)})=(1070\mbox{MeV},50\mbox{MeV})$]
shows that the mass of the ground state is stable for $\sigma_0 > 60$\,MeV and it decreased towards $m_0^{(2)}$ as $\sigma_0$ is decreased from $60$\,MeV.
On the other hand, the right panel [$(1210\mbox{MeV},1050\mbox{MeV})$] shows that all the masses are stable against the change or $\sigma_0$.

Since $\sigma_0$ is an order parameter of chiral symmetry, 
Figs.~\ref{mass_1} and \ref{mass_2} show 
that nucleon masses are degenerated to chiral invariant masses 
when thechiral symmetry is restored in e.g., high temperature and/or density.

\section{summary and discussions}
\label{sec:summary}

We introduced two types of nucleons belonging to $[(\bf{2},\bf{1})\oplus(\bf{1},\bf{2})]$ and $[(\bf{2},\bf{3})\oplus(\bf{3},\bf{2})]$ representations of the chiral ${\rm SU(2)_L}\otimes{\rm SU(2)_R}$ group together with their parity partners. 
We constructed  an effective chiral Lagrangian based on the parity doublet structure. 
We fitted model parameters to the masses, decay widths and axial-charges of $N(939)$,$N(1440)$,$N(1535)$ and $N(1650)$. Our results show that there are five groups of solutions which are separated by 
chiral inavariant masses and mixing structure of nucleons. 
In Group 1, both the chiral invariant masses, $m_0^{(1)}$ for $[(\bf{2},\bf{1})\oplus(\bf{1},\bf{2})]$ and $m_0^{(2)}$ for $[(\bf{2},\bf{3})\oplus(\bf{3},\bf{2})]$, are small as seen from Fig.~\ref{chiralinvariantmass}.
In this group, the ground state $N(939)$ is dominated by   $[(\bf{2},\bf{3})\oplus(\bf{3},\bf{2})]$ component and its chiral partner is $N(1440)$.
In Group 5, on the other hand, the dominant component of $N(939)$ is $[(\bf{2},\bf{1})\oplus(\bf{1},\bf{2})]$, whose chiral invariant mass is about $1000$\,MeV, 
and the chiral partner of $N(939)$ is a mixture of negative parity nucleons.

We gave predictions off-diagonal elements of axial-charge matrix called transition axial-charges,
which shows that some groups can be excluded when some of them are determined by e.g., lattice analysis in future.
We also study the change of nucleon masses when the VEV of $\sigma$, $\sigma_0$, is changed.
In Groups 1-4,
all nucleons masses are decreased with decreasing 
$\sigma_0$ if two chiral invariant masses are smaller than mass of $N(939)$ 
as shown in Fig.~\ref{mass_1}.
In Group 5, on the other hand, the behavior depends on the value of $m_0^{(2)}$:
For small $m_0^{(2)}$ the mass of $N(939)$ is stable for $\sigma_0 > 60$\,MeV and it decreases toward $m_0^{(2)}$, while for large $m_0^{(2)}$ it is stable for all $\sigma_0$.
This seems consistent with  the lattice analysis  in Ref.~\cite{Aarts:2015mma,Aarts:2017rrl}, which shows that, with increasing temperature,  the mass of the 
positive parity nucleon mass is stable, while that of the negative parity nucleon mass decreased.

The chiral representation of $[(\bf{2},\bf{3})\oplus(\bf{3},\bf{2})]$ includes $\Delta$ baryon in addition to nucleons.
For studying $\Delta$, we need to include $[(\bf{4},\bf{1})\oplus(\bf{1},\bf{4})]$ representations which do not include nucleons.
It is interesting to study $\Delta$ baryons by constructing a model including  $[(\bf{4},\bf{1})\oplus(\bf{1},\bf{4})]$ representations.
[Study of $\Delta$ baryon based on the parity doublet structure are done in e.g., 
Refs.~\cite{Jido:1999hd,Jido2,Nagata,Takeda:2017mrm,Bicudo:2016eeu}.]

We can extend  the model to three flavor case based on the SU(3)$_{\rm L}\otimes$SU(3)$_{\rm R}$ symmetry to study hyperons as done in Refs.~\cite{Nemoto1,Chen:2008qv,Chen:2009sf,%
Chen:2010ba,Chen:2011rh,Dmitrasinovic1,Nishihara:2015fka,Olbrich:2015gln,%
Olbrich:2017fsd}. 
The parity doublet structure can also be extended to the baryons including heavy quarks as done in Refs.~\cite{Ma:2015lba,Ma:2015cfa,Ma:2017nik,Kawakami:2018olq}.

It is interesting to construct nuclear matter in the present model and the density dependences of effective masses of nucleons as done in Refs.~\cite{Hatsuda:1988mv,%
Zschiesche:2006zj,Dexheimer:2007tn,Dexheimer:2008cv,Sasaki:2010bp,Gallas:2011qp,%
Steinheimer:2011ea,Dexheimer:2012eu,Benic:2015pia,Motohiro:2015taa,Takeda:2017mrm,%
Suenaga:2017wbb,Takeda:2018ldi,Suenaga:2018kta,Shin:2018axs}.
We leave this for future work.

\acknowledgments
This work was supported partly by JSPS KAKENHI Grant Number JP16K05345.


\begin{thebibliography}{}




\bibitem{DeTar}
  C.~E.~Detar and T.~Kunihiro,
  ``Linear $\sigma$ Model With Parity Doubling,''
  Phys.\ Rev.\ D {\bf 39}, 2805 (1989).
  doi:10.1103/PhysRevD.39.2805

\bibitem{Jido1} 
  D.~Jido, Y.~Nemoto, M.~Oka and A.~Hosaka,
  Nucl.\ Phys.\ A {\bf 671}, 471 (2000)
  doi:10.1016/S0375-9474(99)00844-1
  [hep-ph/9805306].

\bibitem{Jido2} 
  D.~Jido, M.~Oka and A.~Hosaka,
  ``Chiral symmetry of baryons,''
  Prog.\ Theor.\ Phys.\  {\bf 106}, 873 (2001)
  doi:10.1143/PTP.106.873
  [hep-ph/0110005].

\bibitem{Gallas:2009qp} 
  S.~Gallas, F.~Giacosa and D.~H.~Rischke,
  ``Vacuum phenomenology of the chiral partner of the nucleon in a linear sigma model with vector mesons,''
  Phys.\ Rev.\ D {\bf 82}, 014004 (2010)
  doi:10.1103/PhysRevD.82.014004
  [arXiv:0907.5084 [hep-ph]].

\bibitem{Gallas} 
  S.~Gallas and F.~Giacosa,
  ``Mirror versus naive assignment in chiral models for the nucleon,''
  Int.\ J.\ Mod.\ Phys.\ A {\bf 29}, no. 17, 1450098 (2014)
  doi:10.1142/S0217751X14500985
  [arXiv:1308.4817 [hep-ph]].



\bibitem{Nemoto1} 
  Y.~Nemoto, D.~Jido, M.~Oka and A.~Hosaka,
  ``Decays of 1/2- baryons in chiral effective theory,''
  Phys.\ Rev.\ D {\bf 57}, 4124 (1998)
  doi:10.1103/PhysRevD.57.4124
  [hep-ph/9710445].

\bibitem{Chen:2008qv} 
  H.~X.~Chen, V.~Dmitrasinovic, A.~Hosaka, K.~Nagata and S.~L.~Zhu,
  ``Chiral Properties of Baryon Fields with Flavor SU(3) Symmetry,''
  Phys.\ Rev.\ D {\bf 78}, 054021 (2008)
  doi:10.1103/PhysRevD.78.054021
  [arXiv:0806.1997 [hep-ph]].

\bibitem{Dmitrasinovic:2009vp} 
  V.~Dmitrasinovic, A.~Hosaka and K.~Nagata,
  ``Nucleon axial couplings and [(1/2,0) + (0,1/2)] - [(1,1/2) + (1/2,1)] chiral multiplet mixing,''
  Mod.\ Phys.\ Lett.\ A {\bf 25}, 233 (2010)
  doi:10.1142/S0217732310032494
  [arXiv:0912.2372 [hep-ph]].

\bibitem{Dmitrasinovic:2009vy} 
  V.~Dmitrasinovic, A.~Hosaka and K.~Nagata,
  Int.\ J.\ Mod.\ Phys.\ E {\bf 19}, 91 (2010)
  doi:10.1142/S0218301310014650
  [arXiv:0912.2396 [hep-ph]].

\bibitem{Chen:2009sf} 
  H.~X.~Chen, V.~Dmitrasinovic and A.~Hosaka,
  ``Baryon fields with U(L)(3) X U(R)(3) chiral symmetry II: Axial currents of nucleons and hyperons,''
  Phys.\ Rev.\ D {\bf 81}, 054002 (2010)
  doi:10.1103/PhysRevD.81.054002
  [arXiv:0912.4338 [hep-ph]].

\bibitem{Chen:2010ba} 
  H.~X.~Chen, V.~Dmitrasinovic and A.~Hosaka,
  ``Baryon Fields with $U_L(3) times U_R(3)$ Chiral Symmetry III: Interactions with Chiral $(3,\bar{3})+ (\bar{3},3)$ Spinless Mesons,''
  Phys.\ Rev.\ D {\bf 83}, 014015 (2011)
  doi:10.1103/PhysRevD.83.014015
  [arXiv:1009.2422 [hep-ph]].

\bibitem{Chen:2011rh} 
  H.~X.~Chen, V.~Dmitrasinovic and A.~Hosaka,
  ``Baryons with U$_L(3)\times$U$_R(3)$ Chiral Symmetry IV: Interactions with Chiral $(8,1)+(1,8)$ Vector and Axial-vector Mesons and Anomalous Magnetic Moments,''
  Phys.\ Rev.\ C {\bf 85}, 055205 (2012)
  doi:10.1103/PhysRevC.85.055205
  [arXiv:1109.3130 [hep-ph]].

\bibitem{Dmitrasinovic1} 
  V.~Dmitrasinovic, H.~X.~Chen and A.~Hosaka,
  ``Baryon fields with UL(3)×UR(3) chiral symmetry. V. Pion-nucleon and kaon-nucleon Σ terms,''
  Phys.\ Rev.\ C {\bf 93}, no. 6, 065208 (2016).
  doi:10.1103/PhysRevC.93.065208

\bibitem{Nishihara:2015fka} 
  H.~Nishihara and M.~Harada,
  ``Extended Goldberger-Treiman relation in a three-flavor parity doublet model,''
  Phys.\ Rev.\ D {\bf 92}, no. 5, 054022 (2015)
  doi:10.1103/PhysRevD.92.054022
  [arXiv:1506.07956 [hep-ph]].



\bibitem{Olbrich:2015gln} 
  L.~Olbrich, M.~Z\`et\`enyi, F.~Giacosa and D.~H.~Rischke,
  ``Three-flavor chiral effective model with four baryonic multiplets within the mirror assignment,''
  Phys.\ Rev.\ D {\bf 93}, no. 3, 034021 (2016)
  doi:10.1103/PhysRevD.93.034021
  [arXiv:1511.05035 [hep-ph]].



\bibitem{Olbrich:2017fsd} 
  L.~Olbrich, M.~Z\`et\`enyi, F.~Giacosa and D.~H.~Rischke,
  Phys.\ Rev.\ D {\bf 97}, no. 1, 014007 (2018)
  doi:10.1103/PhysRevD.97.014007
  [arXiv:1708.01061 [hep-ph]].


\bibitem{Aarts:2015mma} 
  G.~Aarts, C.~Allton, S.~Hands, B.~J\"ager, C.~Praki and J.~I.~Skullerud,
  ``Nucleons and parity doubling across the deconfinement transition,''
  Phys.\ Rev.\ D {\bf 92}, no. 1, 014503 (2015)
  doi:10.1103/PhysRevD.92.014503
  [arXiv:1502.03603 [hep-lat]].

\bibitem{Aarts:2017rrl} 
  G.~Aarts, C.~Allton, D.~De Boni, S.~Hands, B.~J\"ager, C.~Praki and J.~I.~Skullerud,
  ``Light baryons below and above the deconfinement transition: medium effects and parity doubling,''
  JHEP {\bf 1706}, 034 (2017)
  doi:10.1007/JHEP06(2017)034
  [arXiv:1703.09246 [hep-lat]].




\bibitem{PDGC} Particle Data Group Collaboration
  M.~Tanabashi {\it et al.},
  ``Review of Particle Physics,''
  Phys.\ Rev.\ D {\bf 98}, no. 3, 030001 (2018).
  doi:10.1103/PhysRevD.98.030001


\bibitem{Takahashi:2008fy} 
  T.~T.~Takahashi and T.~Kunihiro,
  Phys.\ Rev.\ D {\bf 78}, 011503 (2008)
  doi:10.1103/PhysRevD.78.011503
  [arXiv:0801.4707 [hep-lat]].


\bibitem{Jido:1999hd} 
  D.~Jido, T.~Hatsuda and T.~Kunihiro,
  ``Chiral symmetry realization for even parity and odd parity baryon resonances,''
  Phys.\ Rev.\ Lett.\  {\bf 84}, 3252 (2000)
  doi:10.1103/PhysRevLett.84.3252
  [hep-ph/9910375].

\bibitem{Nagata} 
  K.~Nagata, A.~Hosaka and V.~Dmitrasinovic,
  ``pi N and pi pi N Couplings of the Delta(1232) and its Chiral Partners,''
  Phys.\ Rev.\ Lett.\  {\bf 101}, 092001 (2008)
  doi:10.1103/PhysRevLett.101.092001
  [arXiv:0804.3185 [hep-ph]].

\bibitem{Takeda:2017mrm} 
  Y.~Takeda, Y.~Kim and M.~Harada,
  ``Catalysis of partial chiral symmetry restoration by $\Delta$ matter,''
  Phys.\ Rev.\ C {\bf 97}, no. 6, 065202 (2018)
  doi:10.1103/PhysRevC.97.065202
  [arXiv:1704.04357 [nucl-th]].

\bibitem{Bicudo:2016eeu} 
  P.~Bicudo, M.~Cardoso, F.~J.~Llanes-Estrada and T.~Van Cauteren,
  ``Mapping chiral symmetry breaking in the excited baryon spectrum,''
  Phys.\ Rev.\ D {\bf 94}, no. 5, 054006 (2016)
  doi:10.1103/PhysRevD.94.054006
  [arXiv:1605.05171 [hep-ph]].



\bibitem{Ma:2015lba} 
  Y.~L.~Ma and M.~Harada,
  ``Doubly heavy baryons with chiral partner structure,''
  Phys.\ Lett.\ B {\bf 748}, 463 (2015)
  doi:10.1016/j.physletb.2015.07.046
  [arXiv:1503.05373 [hep-ph]].

\bibitem{Ma:2015cfa} 
  Y.~L.~Ma and M.~Harada,
  ``Degeneracy of doubly heavy baryons from heavy quark symmetry,''
  Phys.\ Lett.\ B {\bf 754}, 125 (2016)
  doi:10.1016/j.physletb.2016.01.011
  [arXiv:1510.07481 [hep-ph]].

\bibitem{Ma:2017nik} 
  Y.~L.~Ma and M.~Harada,
  ``Chiral partner structure of doubly heavy baryons with heavy quark spin-flavor symmetry,''
  J.\ Phys.\ G {\bf 45}, no. 7, 075006 (2018)
  doi:10.1088/1361-6471/aac86e
  [arXiv:1709.09746 [hep-ph]].

\bibitem{Kawakami:2018olq} 
  Y.~Kawakami and M.~Harada,
  ``Analysis of $\Lambda_c(2595)$, $\Lambda_c(2625)$, $\Lambda_b(5912)$, $\Lambda_b(5920)$ based on a chiral partner structure,''
  Phys.\ Rev.\ D {\bf 97}, no. 11, 114024 (2018)
  doi:10.1103/PhysRevD.97.114024
  [arXiv:1804.04872 [hep-ph]].





\bibitem{Hatsuda:1988mv} 
  T.~Hatsuda and M.~Prakash,
  ``Parity Doubling of the Nucleon and First Order Chiral Transition in Dense Matter,''
  Phys.\ Lett.\ B {\bf 224}, 11 (1989).
  doi:10.1016/0370-2693(89)91040-X

\bibitem{Zschiesche:2006zj} 
  D.~Zschiesche, L.~Tolos, J.~Schaffner-Bielich and R.~D.~Pisarski,
  ``Cold, dense nuclear matter in a SU(2) parity doublet model,''
  Phys.\ Rev.\ C {\bf 75}, 055202 (2007)
  doi:10.1103/PhysRevC.75.055202
  [nucl-th/0608044].

\bibitem{Dexheimer:2007tn} 
  V.~Dexheimer, S.~Schramm and D.~Zschiesche,
  ``Nuclear matter and neutron stars in a parity doublet model,''
  Phys.\ Rev.\ C {\bf 77}, 025803 (2008)
  doi:10.1103/PhysRevC.77.025803
  [arXiv:0710.4192 [nucl-th]].

\bibitem{Dexheimer:2008cv} 
  V.~Dexheimer, G.~Pagliara, L.~Tolos, J.~Schaffner-Bielich and S.~Schramm,
  ``Neutron stars within the SU(2) parity doublet model,''
  Eur.\ Phys.\ J.\ A {\bf 38}, 105 (2008)
  doi:10.1140/epja/i2008-10652-0
  [arXiv:0805.3301 [nucl-th]].

\bibitem{Sasaki:2010bp} 
  C.~Sasaki and I.~Mishustin,
  ``Thermodynamics of dense hadronic matter in a parity doublet model,''
  Phys.\ Rev.\ C {\bf 82}, 035204 (2010)
  doi:10.1103/PhysRevC.82.035204
  [arXiv:1005.4811 [hep-ph]].

\bibitem{Gallas:2011qp} 
  S.~Gallas, F.~Giacosa and G.~Pagliara,
  ``Nuclear matter within a dilatation-invariant parity doublet model: the role of the tetraquark at nonzero density,''
  Nucl.\ Phys.\ A {\bf 872}, 13 (2011)
  doi:10.1016/j.nuclphysa.2011.09.008
  [arXiv:1105.5003 [hep-ph]].

\bibitem{Steinheimer:2011ea} 
  J.~Steinheimer, S.~Schramm and H.~Stocker,
  ``The hadronic SU(3) Parity Doublet Model for Dense Matter, its extension to quarks and the strange equation of state,''
  Phys.\ Rev.\ C {\bf 84}, 045208 (2011)
  doi:10.1103/PhysRevC.84.045208
  [arXiv:1108.2596 [hep-ph]].

\bibitem{Dexheimer:2012eu} 
  V.~Dexheimer, J.~Steinheimer, R.~Negreiros and S.~Schramm,
  Phys.\ Rev.\ C {\bf 87}, no. 1, 015804 (2013)
  doi:10.1103/PhysRevC.87.015804
  [arXiv:1206.3086 [astro-ph.HE]].


\bibitem{Benic:2015pia} 
  S.~Benic, I.~Mishustin and C.~Sasaki,
  Phys.\ Rev.\ D {\bf 91}, no. 12, 125034 (2015)
  doi:10.1103/PhysRevD.91.125034
  [arXiv:1502.05969 [hep-ph]].


\bibitem{Motohiro:2015taa} 
  Y.~Motohiro, Y.~Kim and M.~Harada,
  ``Asymmetric nuclear matter in a parity doublet model with hidden local symmetry,''
  Phys.\ Rev.\ C {\bf 92}, no. 2, 025201 (2015)
  Erratum: [Phys.\ Rev.\ C {\bf 95}, no. 5, 059903 (2017)]
  doi:10.1103/PhysRevC.92.025201, 10.1103/PhysRevC.95.059903
  [arXiv:1505.00988 [nucl-th]].

\bibitem{Suenaga:2017wbb} 
  D.~Suenaga,
  Phys.\ Rev.\ C {\bf 97}, no. 4, 045203 (2018)
  doi:10.1103/PhysRevC.97.045203
  [arXiv:1704.03630 [nucl-th]].

\bibitem{Takeda:2018ldi} 
  Y.~Takeda, H.~Abuki and M.~Harada,
  ``Novel dual chiral density wave in nuclear matter based on a parity doublet structure,''
  Phys.\ Rev.\ D {\bf 97}, no. 9, 094032 (2018)
  doi:10.1103/PhysRevD.97.094032
  [arXiv:1803.06779 [hep-ph]].

\bibitem{Suenaga:2018kta} 
  D.~Suenaga,
  ``Spectral function for $\bar{D}_{0}^{\ast}$ $(0^+)$ meson in isospin asymmetric nuclear matter with chiral partner structure,''
  arXiv:1805.01709 [nucl-th].

\bibitem{Shin:2018axs} 
  I.~J.~Shin, W.~G.~Paeng, M.~Harada and Y.~Kim,
  ``Nuclear structure in Parity Doublet Model,''
  arXiv:1805.03402 [nucl-th].






\end{thebibliography}
\end{document}